\documentclass[twocolumn,twocolappendix,trackchanges]{aastex702}

\usepackage{threeparttable} 
\usepackage{amsmath}
\usepackage[absolute,overlay]{textpos}
\usepackage{xcolor}
\graphicspath{{./}{figures/}}

\shorttitle{}
\shortauthors{Han et al.}

\begin{document}

\title{AGN Host Galaxies in COSMOS-Web: Linking Structural Evolution and Star Formation Quenching up to $z=5$
}

\author[0009-0004-0902-0611]{Xiao Han}
\email{hanxiaoastro@whu.edu.cn}
\affiliation{School of Physics and Technology, Wuhan University, Wuhan 430072, China}

\author[0000-0002-8434-880X]{Lilan Yang}
\email[show]{yanglilan@hunnu.edu.cn}
\affiliation{Department of Physics, School of Physics and Electronics, Hunan Normal University, Changsha 410081, People’s Republic of China}
\affiliation{Laboratory for Multiwavelength Astrophysics, School of Physics and Astronomy, Rochester Institute of Technology, 84 Lomb Memorial Drive, Rochester, NY 14623, USA}

\author[0009-0005-9497-7289]{T. Kelly}
\email{tsk4215@g.rit.edu}
\affiliation{Laboratory for Multiwavelength Astrophysics, School of Physics and Astronomy, Rochester Institute of Technology, 84 Lomb Memorial Drive, Rochester, NY 14623, USA}

\author[0000-0001-6919-1237]{Matthew A. Malkan}
\email{malkan@astro.ucla.edu}
\affiliation{Department of Physics and Astronomy, University of California, 475 Portola Plaza, Los Angeles, CA 90095, USA}

\author[0000-0001-6947-5846]{Luis C. Ho}
\email{lho.pku@gmail.com}
\affiliation{Kavli Institute for Astronomy and Astrophysics, Peking University, Beijing 100871, China}
\affiliation{Department of Astronomy, School of Physics, Peking University, Beijing 100871, China}

\author[0000-0002-0000-6977]{John D. Silverman}
\email{john.silverman@ipmu.jp}
\affiliation{Kavli Institute for the Physics and Mathematics of the Universe (WPI), The University of Tokyo Institutes for Advanced Study, The University of Tokyo, 5-1-5 Kashiwanoha, Kashiwa, Chiba 277-8583, Japan}
\affiliation{Department of Astronomy, Graduate School of Science, The University of Tokyo, 7-3-1 Hongo, Bunkyo-ku, Tokyo 113-0033, Japan}
\affiliation{Center for Data-Driven Discovery, Kavli IPMU (WPI), UTIAS, The University of Tokyo, 5-1-5 Kashiwanoha, Kashiwa, Chiba 277-8583, Japan}
\affiliation{Center for Astrophysical Sciences, Department of Physics and Astronomy, Johns Hopkins University, 3400 N. Charles Street, Baltimore, MD 21218, USA}

\author[0000-0001-7634-1547]{Zi-Jian Li}
\affiliation{School of Physics and Technology, Wuhan University, Wuhan 430072, China}
\email{zjli@nao.cas.cn}

\author[0000-0002-9382-9832]{Andreas L. Faisst}
\email{afaisst@caltech.edu}
\affiliation{Caltech/IPAC, MS 314-6, 1200 E. California Blvd. Pasadena, CA 91125, USA}

\author[0000-0003-4569-2285]{Andrew J. Battisti}
\email{andrew.battisti@anu.edu.au}
\affiliation{International Centre for Radio Astronomy Research, University of Western Australia, 35 Stirling Hwy, Crawley, WA 6009, Australia}
\affiliation{Research School of Astronomy and Astrophysics, Australian National University, Cotter Road, Weston Creek, ACT 2611, Australia}

\author[0000-0001-9187-3605]{Jeyhan S. Kartaltepe}
\email{jeyhan@astro.rit.edu}
\affiliation{Laboratory for Multiwavelength Astrophysics, School of Physics and Astronomy, Rochester Institute of Technology, 84 Lomb Memorial Drive, Rochester, NY 14623, USA}

\author[0000-0002-3473-6716]{Lauro Moscardini}
\email{lauro.moscardini@unibo.it}
\affiliation{University of Bologna - Department of Physics and Astronomy “Augusto Righi” (DIFA), Via Gobetti 93/2, I-40129 Bologna, Italy}
\affiliation{INAF- Osservatorio di Astrofisica e Scienza dello Spazio, Via Gobetti 93/3, I-40129 Bologna, Italy}
\affiliation{INFN- Sezione di Bologna, Viale Berti Pichat 6/2, I-40127 Bologna, Italy}

\author[0000-0003-2716-8332]{Rasha M. Samir}
\affiliation{Department of Astronomy, National Research Institute of Astronomy and Geophysics (NRIAG), Cairo, 11421, Egypt}
\email{rasha.samir@nriag.sci.eg}

\author[0000-0002-0322-6131]{Ronaldo Laishram}
\email{ronaldo.laishram@nao.ac.jp}
\affiliation{National Astronomical Observatory of Japan, 2-21-1 Osawa, Mitaka, Tokyo 181-8588, Japan}

\author[0000-0002-2504-2421]{Tao Wang}
\affiliation{School of Astronomy and Space Science, Nanjing University, Nanjing, Jiangsu 210093, China}
\affiliation{Key Laboratory of Modern Astronomy and Astrophysics, Nanjing University, Ministry of Education, Nanjing 210093, China}
\email{taowang@nju.edu.cn}

\author[0009-0006-8707-023X]{Yiyang Zhang}
\email{yiyang_zhang@whu.edu.cn}
\affiliation{School of Physics and Technology, Wuhan University, Wuhan 430072, China}

\author[0000-0002-0930-6466]{Caitlin M. Casey}
\email{cmcasey@ucsb.edu}
\affiliation{Department of Physics, University of California, Santa Barbara, Santa Barbara, CA 93106, USA}
\affiliation{Cosmic Dawn Center (DAWN), Denmark}

\author[0000-0002-9489-7765]{Henry Joy McCracken}
\email{hjmcc@iap.fr}
\affiliation{Institut d’Astrophysique de Paris, UMR 7095, CNRS, and Sorbonne Université, 98 bis boulevard Arago, F-75014 Paris, France}

\author[0000-0002-4485-8549]{Jason Rhodes}
\affiliation{Jet Propulsion Laboratory, California Institute of Technology, 4800 Oak Grove Drive, Pasadena, CA 91001, USA}
\email{}

\author[0000-0002-4271-0364]{Brant E. Robertson}
\affiliation{Department of Astronomy and Astrophysics, University of California, Santa Cruz, 1156 High Street, Santa Cruz, CA 95064, USA}
\email{brant@ucsc.edu}

\author[0000-0002-3560-8599]{Maximilien Franco}
\email{maximilien.franco@iap.fr}
\affiliation{Institut d’Astrophysique de Paris, UMR 7095, CNRS, Sorbonne Université, 98 bis boulevard Arago, 75014 Paris, France}

\author[0000-0001-9773-7479]{Daizhong Liu}
\email{dzliu@pmo.ac.cn}
\affiliation{Purple Mountain Observatory, Chinese Academy of Sciences, 10 Yuanhua Road, Nanjing 210023, China}

\author[0000-0002-7087-0701]{Marko Shuntov}
\email{marko.shuntov@nbi.ku.dk}
\affiliation{Cosmic Dawn Center (DAWN), Denmark} 
\affiliation{Niels Bohr Institute, University of Copenhagen, Jagtvej 128, DK-2200, Copenhagen, Denmark}
\affiliation{University of Geneva, 24 rue du Général-Dufour, 1211 Genève 4, Switzerland}

\author[0000-0001-8917-2148]{Xuheng Ding}
\email{dingxh@whu.edu.cn}
\affiliation{School of Physics and Technology, Wuhan University, Wuhan 430072, China}

\begin{abstract}
The role of active galactic nuclei (AGNs) in the suppression of star formation and structural evolution of galaxies during their transition from star-forming to quiescent systems remains unclear. We analyze 3,955 AGNs selected through five multiwavelength diagnostics at $0<z<5$. Using high-resolution JWST imaging from the COSMOS-Web survey and archival multi-wavelength photometry, we investigate the rest-frame optical size-mass distribution of AGN host galaxies and compare it with those of star-forming galaxies (SFGs) and quiescent galaxies (QGs). We find that the positions of AGN hosts in the size-mass plane depend on both redshift and stellar mass. At $z\gtrsim2$, massive AGN hosts ($M_\star/M_\odot>10^{10.5}$) generally exhibit sizes comparable to those of SFGs despite suppressed star formation. Toward lower redshifts, they increasingly approach the QG size-mass relation, accompanied by declining specific star formation rates (sSFRs) and increasing S\'ersic indices. In contrast, lower-mass AGN hosts remain close to the SFG relation across the redshift range. Overall, AGN hosts exhibit relatively mild size evolution, following $R_e\propto(1+z)^{-0.6}$ over $0<z<4$, shallower than that of typical SFGs and QGs. We further examine the relationship between star formation activity and structural compactness in the $\Delta\log\mathrm{SFR}_{\mathrm{MS}}$-$\Delta\Sigma_e$ plane. At $z<2.2$, AGN hosts predominantly occupy the compact star-forming and quiescent regions, broadly consistent with the compaction-quenching scenario. At $z>2.2$, however, some AGN hosts exhibit suppressed star formation while retaining relatively extended structures. These findings suggest that some AGN hosts may follow alternative evolutionary pathways in which star formation suppression precedes substantial structural compaction.

\end{abstract}


\section{Introduction}
\label{introduction}

The structural properties of galaxies are potential tracers of their assembly history, reflecting the interplay between star formation, gas accretion, mergers, and feedback processes \citep{2014ARA&A..52..291C, 2015ARA&A..53...51S}. Observations have established distinct size--mass relations for star-forming galaxies (SFGs) and quiescent galaxies (QGs) across a broad redshift range \citep{2014ApJ...788...28V, 2025ApJS..281...68Y,2026ApJ..1002L..40C}. At fixed stellar mass, QGs generally exhibit smaller effective radius and higher S\'ersic index than SFGs, whereas SFGs tend to have more extended, disk-like structures and higher specific star formation rates (sSFRs; $\mathrm{SFR}/M_\star$). These differences suggest that the transition from star-forming to quiescent systems involves changes in galaxy structure as well as star formation activity. Identifying the physical mechanisms that drive this transformation is a central problem in galaxy evolution.

Active galactic nucleus (AGN) feedback has been proposed as a potential mechanism for regulating star formation, particularly in massive galaxies. Radiation, AGN-driven winds, and relativistic jets can transfer energy and momentum to the interstellar and circumgalactic media, potentially removing or preventing the cooling of gas required for star formation \citep{2025arXiv250201724D,2012ARA&A..50..455F,2014ARA&A..52..589H}. Meanwhile, there is some connection between AGN activity and galaxy structure \citep{2023ApJ...946L..14K}. For example, studies over $0.2 \lesssim z \lesssim 3$ have found that AGN hosts generally occupy an intermediate region between the size--mass relations of SFGs and QGs \citep{2019ApJ...887L...5S,2021ApJ...918...22L,2024MNRAS.527.4690L,2025ApJ...994..265V}. Using COSMOS-Web observations, \citet{2024ApJ...962...93Z} found that X-ray-selected broad-line AGN hosts at $0.35 \lesssim z \lesssim 3.5$ broadly follow the size--mass relations of inactive galaxies, with star-forming AGN hosts showing slightly smaller sizes than their non-AGN counterparts. At higher redshifts ($4 \lesssim z \lesssim 7$), \citet{2026ApJ...999...70L} also found that quasar host galaxies are systematically more compact than SFGs at similar stellar masses.

These findings suggest that AGN activity may be associated with both the regulation of star formation and the structural evolution of galaxies. However, the physical mechanism linking these processes, as well as their causal relationships and relative timing, remain poorly understood. In particular, it remains unclear whether AGN activity precedes, accompanies, or follows the structural transformation and quenching of its host galaxy. Therefore, studying the structural properties and star formation activities of AGN host galaxies is important to investigate the impact of AGN on galaxy evolution.

\begin{figure*}
  \centering
  \includegraphics[width=\textwidth]{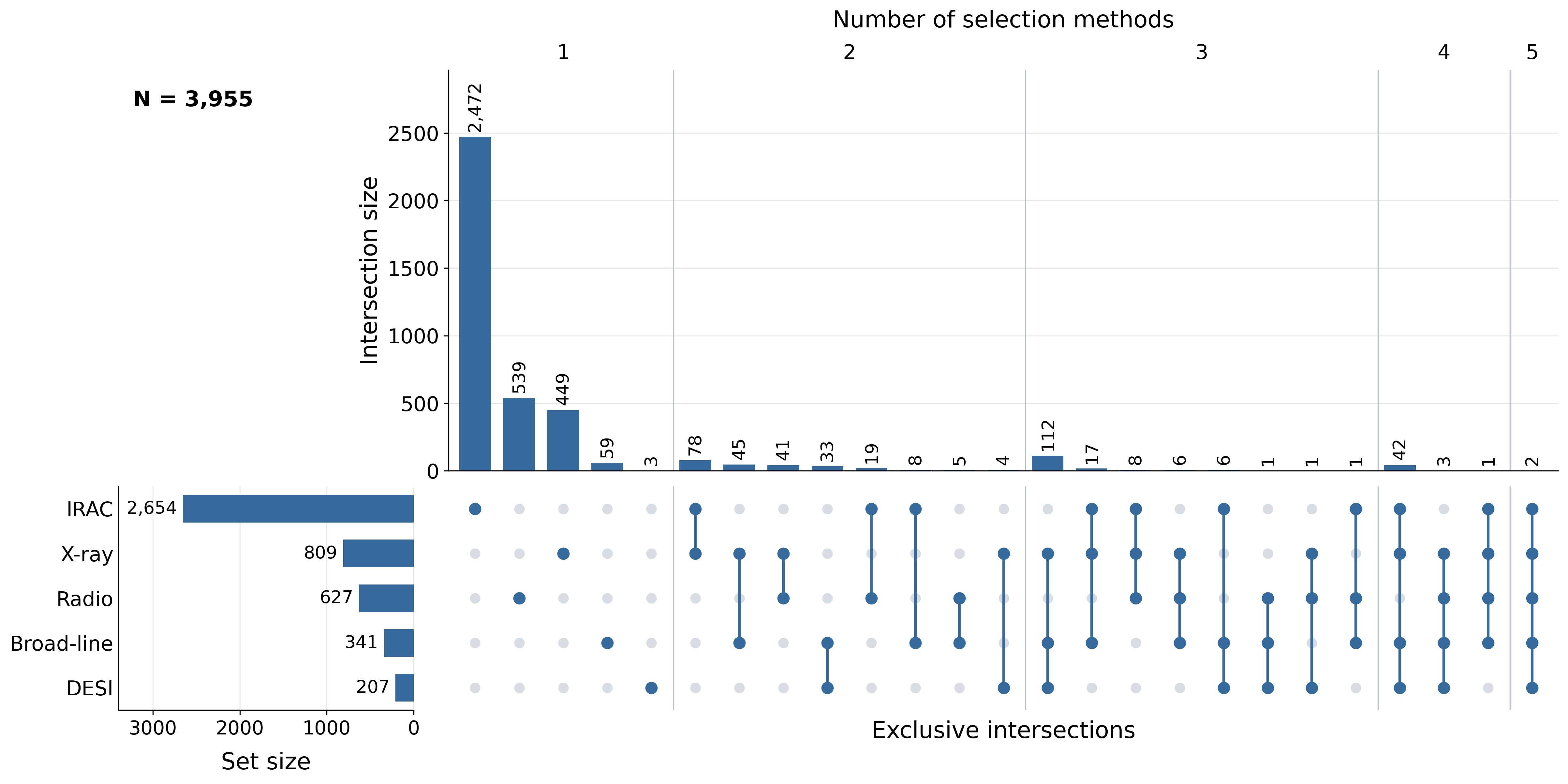}
  \caption{Overlap among the five AGN selection methods (IRAC, X-ray, radio, broad-line, and DESI spectroscopy) for the final sample of 3,955 unique sources. The horizontal bars on the left show the total number of AGNs selected by each method. Each column in the bottom matrix represents one combination of selections, with blue dots indicating the selections included in that combination. The vertical bar above each column shows the number of sources selected by the corresponding combination.}

  \label{selection_upset}
\end{figure*}

One possible physical interpretation is provided by the compaction--quenching framework. In this scenario, intense gas inflows---driven by processes such as mergers, counter-rotating streams \citep{2023NatAs...7..463R}, or violent disk instabilities---lead to dissipative contraction, build a dense central stellar core, and trigger central star formation \citep{2015MNRAS.450.2327Z,2016MNRAS.458..242T,2017ApJ...840...47B, 2026ApJ..1009..106H}. Following compaction, rapid central gas depletion, together with reduced gas inflow or replenishment, may initiate inside-out quenching. Observationally, massive compact SFGs at $0.5\lesssim z\lesssim3$ have shown an elevated incidence of AGN, and the probability of hosting an AGN increases with compactness among the SFGs \citep{2013ApJ...765..104B,2014ApJ...791...52B,2017ApJ...846..112K, 2020PASP..132e4101G,2022MNRAS.515.4860A}. AGN feedback could further promote quenching by removing or heating cold gas and suppressing subsequent cooling. In this framework, central gas inflows, black hole accretion, and galaxy compaction may be closely linked during the transition from SFGs to QGs.

Interpreting AGN host properties is complicated by the fact that different AGN selection techniques trace different emission components, with varying sensitivity to accretion rate and obscuration. Consequently, the resulting samples may exhibit systematically different host galaxy structures. Previous studies have found that radio-selected AGNs predominantly reside in massive, red, bulge-dominated galaxies, whereas mid-infrared selected AGNs are more often associated with bluer, disk-dominated hosts; with X-ray selected AGNs occupying an intermediate range in host color and morphology \citep{2009ApJ...696..891H, 2010AJ....140..533G}. X-ray selected AGNs tend to inhabit more massive and redder hosts \citep{2005ApJ...621..587C}, while IR selected AGNs are preferentially found in less massive galaxies with enhanced star formation \citep{2022ApJ...925...74J}. IRAC-selected AGNs preferentially trace luminous, heavily obscured systems \citep{2026ApJ...997..366K}, whose host galaxies show higher incidences of irregular, asymmetric, and disturbed morphologies than those of X-ray-selected AGNs \citep{2018ApJ...853...63D}. 

Despite these advances, previous multiwavelength studies have generally been limited to one or two AGN selection methods and, in many cases, to relatively small samples or restricted redshift ranges. A large, multiwavelength-selected sample with uniform structural measurements is therefore needed to investigate the dependence of host galaxy structure on stellar mass, redshift, and AGN selection method.

The COSMOS-Web survey provides an opportunity to address these questions by combining deep, high-resolution JWST imaging with extensive multiwavelength observations. In this work, we analyze 3,955 AGNs selected through mid-infrared, X-ray, radio-excess selection, as well as broad-line and DESI spectroscopic classifications (T. Kelly et al., in preparation). Using AGN--host image decomposition and spectral energy distribution (SED) fitting, we investigate the rest-frame optical size--mass relations, Sérsic indices, sSFR, and stellar mass surface densities of AGN host galaxies. Our primary objectives are to determine how these properties depend on stellar mass and redshift and to compare host galaxy structures across different AGN selection methods. Finally, we discuss the implications of our results for the relationship between AGN activity, structural compaction, and star formation quenching.

The paper is organized as follows. Section \ref{data} describes the COSMOS-Web data products and the parent multi-wavelength AGN catalog. Section \ref{method} presents the construction of the AGN sample and the measurements of host galaxy physical and structural properties. Section \ref{results} presents the size--mass relation of AGN host galaxies and their redshift evolution. Section \ref{discussion} examines the relation between star formation and galaxy compactness, compares host galaxy properties across different AGN selection methods, and discusses the possible role of AGN in the evolution from SFGs to QGs. Section \ref{summary} summarizes our main conclusions. Throughout this work, we adopt a flat $\Lambda$CDM cosmology with $H_0=70\ {\rm km\ s^{-1}\ Mpc^{-1}}$, $\Omega_{\rm m}=0.3$, and $\Omega_{\Lambda}=0.7$, and quote magnitudes in the AB system \citep{1983ApJ...266..713O}. Unless otherwise stated, stellar masses refer to the fiducial \texttt{CIGALE} estimates assuming a \citet{2003PASP..115..763C} initial mass function (IMF); the independent \texttt{BAGPIPES} validation instead adopts a \citet{2001MNRAS.322..231K} IMF.

\section{Data and Catalogs} 
\label{data}

\subsection{COSMOS-Web Data}
\label{COSMOS-Web data}

COSMOS-Web is a 255\,hr JWST Cycle~1 Treasury Program (GO~1727; PIs: Kartaltepe and Casey) designed to bridge the gap between narrow, deep fields and wider, shallower surveys \citep{2023ApJ...954...31C}. The survey covers a contiguous area of approximately $0.54\,\mathrm{deg}^2$ in the central COSMOS field with NIRCam imaging in F115W, F150W, F277W, and F444W, complemented by approximately $0.19\,\mathrm{deg}^{2}$ of non-contiguous parallel MIRI/F770W imaging. 

\subsection{Photometric Catalog}
\label{photometric catalog}

We adopt the multi-wavelength photometry from COSMOS2025 \citep{2025A&A...704A.339S}, a galaxy catalog constructed for the COSMOS-Web footprint by combining JWST observations with ancillary imaging from HST, CFHT, Subaru, VISTA, and Spitzer. The catalog provides photometry, morphological measurements, photometric redshifts, and physical parameters for more than 700,000 NIRCam-detected sources. Sources are detected in a combined four-band NIRCam image using a two-stage \textit{hot+cold} procedure implemented with \texttt{SEP}, a Python implementation of the Source Extractor algorithms \citep{1996A&AS..117..393B,Barbary2016}. The cold mode is optimized for bright and extended objects, while the hot mode recovers faint, isolated sources. The resulting detection catalog provides positional and morphological priors for \texttt{SourceXtractor++} (\texttt{SE++}), which fits PSF-convolved parametric profiles to neighboring sources in the native-resolution images. Structural parameters and total fluxes are first fitted simultaneously in the four NIRCam bands, followed by forced photometry in the remaining bands. For our analysis, we adopt the resulting model-based total fluxes and calibrated uncertainties in 37 bands spanning observed wavelengths of  approximately $0.3$--$8\,\mu{\rm m}$.

\begin{figure}
    \centering
    \includegraphics[width=\columnwidth]{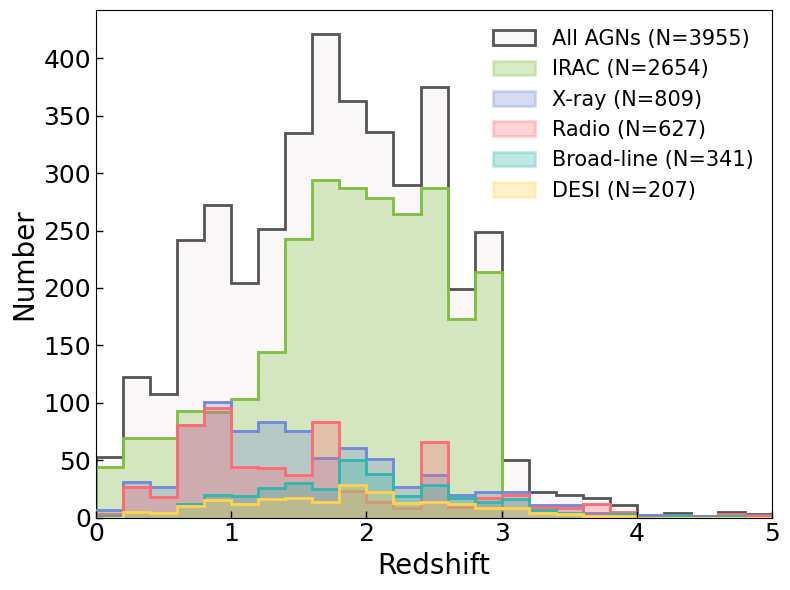}
    \caption{
    Redshift distributions of the AGNs at $0<z<5$. The gray filled histogram with the dark outline represents all 3,955 AGNs, while the colored histograms show the IRAC-, X-ray-, radio-, broad-line-, and DESI-selected subsamples. The number of objects in each displayed subsample is given in the legend.
    }
    \label{redshift_distribution}
\end{figure}

Photometric redshifts in COSMOS2025 are estimated through multiband template fitting with \texttt{LePHARE} \citep{2002MNRAS.329..355A,2006A&A...457..841I}. For each source,
galaxy, stellar, and AGN template libraries are fitted in parallel, and their best-fit statistics are combined with color, compactness, aperture-flux ratio, and X-ray information to classify sources as galaxies, stars, or AGNs. The galaxy template library is based on the stellar population models of \citet{2003MNRAS.344.1000B}, spanning a range of star formation histories (SFHs), metallicities, ages, nebular emission, dust attenuation, and intergalactic absorption. The stellar template library follows \citet{2022A&A...667A..65K} and includes brown-dwarf templates to reduce degeneracies with high redshift galaxies. The AGN templates adopt the hybrid galaxy--AGN models of \citet{2011ApJ...742...61S}. We note that this AGN classification is not exhaustive and primarily identifies luminous, relatively unobscured AGNs whose nuclear emission can significantly alter the broadband SED.

The fiducial photometric redshift is defined as the median of the redshift probability distribution, with uncertainties given by the central 68\% interval. Comparison with approximately 12,000 high-confidence spectroscopic redshifts extending to $z=8$ yields $\sigma_{\rm MAD}=0.012$ for sources with $m_{\rm F444W}<28$, increasing to 0.030 in the faintest magnitude bin ( $26<m_{\rm F444W}<28$). An independent comparison at $z\gtrsim6.5$ yields $\sigma_{\rm MAD}=0.038$. The catastrophic failure rate of photometric redshift estimates is low, with only 1.98\% of sources classified as outliers in the spectroscopic validation sample with $m_{\rm F444W}<28$ \citep{2025A&A...704A.339S}.

COSMOS2025 also provides an independent set of physical parameters derived with \texttt{CIGALE} \citep{2019A&A...622A.103B}, with redshifts fixed to the \texttt{LePHARE} solution and non-parametric SFHs \citep{2025A&A...704A.339S}. The modeling attempts to fit dust attenuation, and includes possible contributions from  stellar and nebular emission, as well as an AGN contribution when necessary. The resulting fitted parameters include stellar masses, star formation rates (SFR), dust attenuation, stellar metallicities, and time-resolved SFHs.

\subsection{Morphological Catalog}
\label{morphological catalog}

We adopt the structural measurements of AGN host galaxies from the COSMOS-Web multiwavelength morphology catalog \citep{2026arXiv260614869Y}, which performs two-dimensional surface-brightness decomposition using \texttt{GALIGHT} \citep{2020ApJ...888...37D}. The image fitting was performed independently in the four JWST/NIRCam bands (F115W, F150W, F277W, and F444W), using native resolution mosaics with a pixel scale of 30 mas. Each source was modeled with three possible configurations: a single S\'ersic profile, a two-component bulge--disk model, and a central point source combined with an extended S\'ersic component. The fits incorporated the band-specific point spread function (PSF), noise map, and segmentation information to account for neighboring sources. For our AGN sample, we adopt the point source plus S\'ersic decomposition, in which the point source represents unresolved nuclear emission and the S\'ersic component represents the extended host emission. This simplified model may provide poor fits to galaxy profiles with substantial deviations, such as bars, spiral arms, dust lanes and brightness clumps. Such cases are flagged in the morphology catalog by large $\chi^2$ values.

\subsection{AGN Catalog}
\label{AGN catalog}

\begin{figure}
  \centering
  \includegraphics[width=\columnwidth]{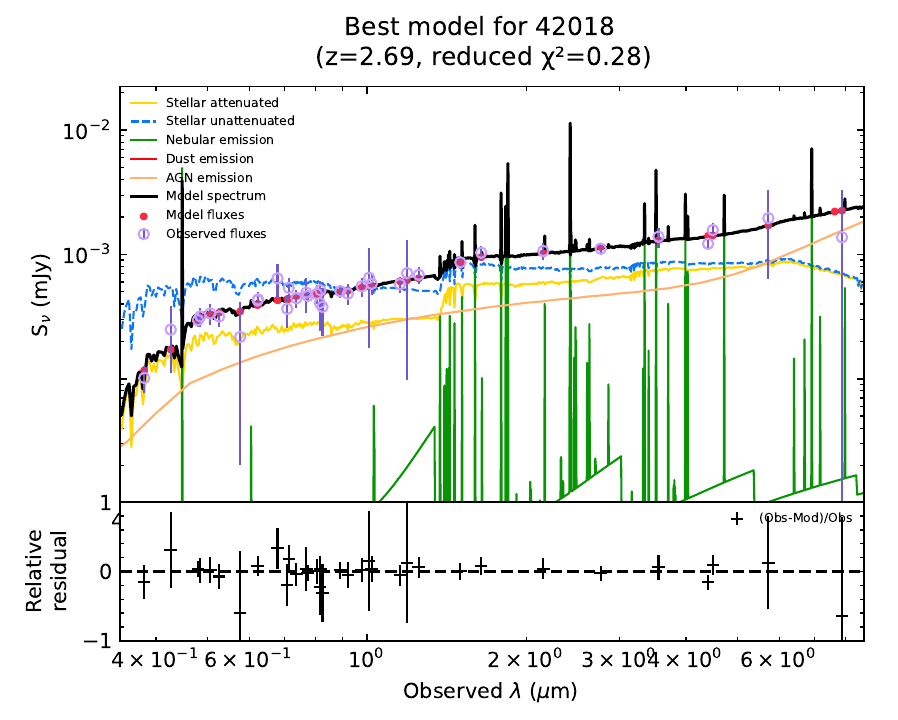}
  \caption{Observed SED of source ID 42018 and its best-fitting \texttt{CIGALE} model at $z=2.69$. In the upper panel, the open purple circles show the observed photometric flux densities with their $1\sigma$ uncertainties, while the filled red circles represent the corresponding model fluxes. The black curve denotes the total best-fitting model spectrum. The colored curves show the attenuated stellar emission (yellow), unattenuated stellar emission (blue dashed), nebular emission (green), dust emission (red), and AGN emission (orange). The lower panel presents the relative residuals, $(S_{\nu,\mathrm{obs}}-S_{\nu,\mathrm{mod}})/S_{\nu,\mathrm{obs}}$. The best-fitting model has a reduced $\chi^2$ value of $\chi_{\nu}^{2}=0.28$.
  }
  \label{42018_best_model}
\end{figure}

We adopt the COSMOS AGN multiwavelength emission tabulation (AMMIT) Project (T. Kelly et al., in preparation). The catalog compiles AGNs identified through five complementary selection methods spanning multiple wavelengths, including X-ray, UV, optical, infrared, and radio. It contains 13,891 unique AGN candidates across the COSMOS field, of which 1,804 are identified by more than one selection method. The AGNs extend to redshifts of $z\sim12$, although the vast majority (96.7\%) lie at $z<3$.

Specifically, X-ray selected AGNs are defined as sources with intrinsic 0.5--10 keV luminosities $L_{\rm X} > 10^{42}\ {\rm erg\ s^{-1}}$, based on the Chandra-COSMOS Legacy Survey \citep{2016ApJ...819...62C,2016ApJ...817...34M}. Mid-infrared AGNs are selected using the observed-frame Spitzer/IRAC color criteria of \citet{2012ApJ...748..142D}, based on COSMOS2020 photometry \citep{2022ApJS..258...11W}. Since the observed-frame IRAC bands probe shorter rest-frame wavelengths at higher redshifts, the selection is restricted to $z<3$ to minimize contamination from SFGs (T. Kelly et al., in preparation). Radio-selected AGNs are identified using the radio-excess classification \citep{1996ApJ...473..130R, 2017A&A...602A...3D}, based on observations from the VLA-COSMOS 3 GHz Large Project \citep{2017A&A...602A...1S,2017A&A...602A...6S}. Sources with rest-frame 1.4 GHz radio luminosities exceeding the redshift-dependent threshold relative to their infrared-derived SFRs are classified as AGNs. 

Broad-line AGNs are identified through the presence of broad permitted emission lines in their spectra, as listed in the COSMOS Spectroscopic Redshift Compilation \citep{2026ApJS..282....6K}. Additional spectroscopically selected AGNs are drawn from the DESI DR1 and DESI-COSMOS surveys \citep{2026AJ....171..285D, 2026AJ....171...71R}. The DESI AGN selection includes sources identified through broad emission lines, high-excitation narrow emission lines, AGN-sensitive emission line diagnostics, and non-stellar continuum features.

\begin{figure*}
  \centering
  \includegraphics[width=\textwidth]{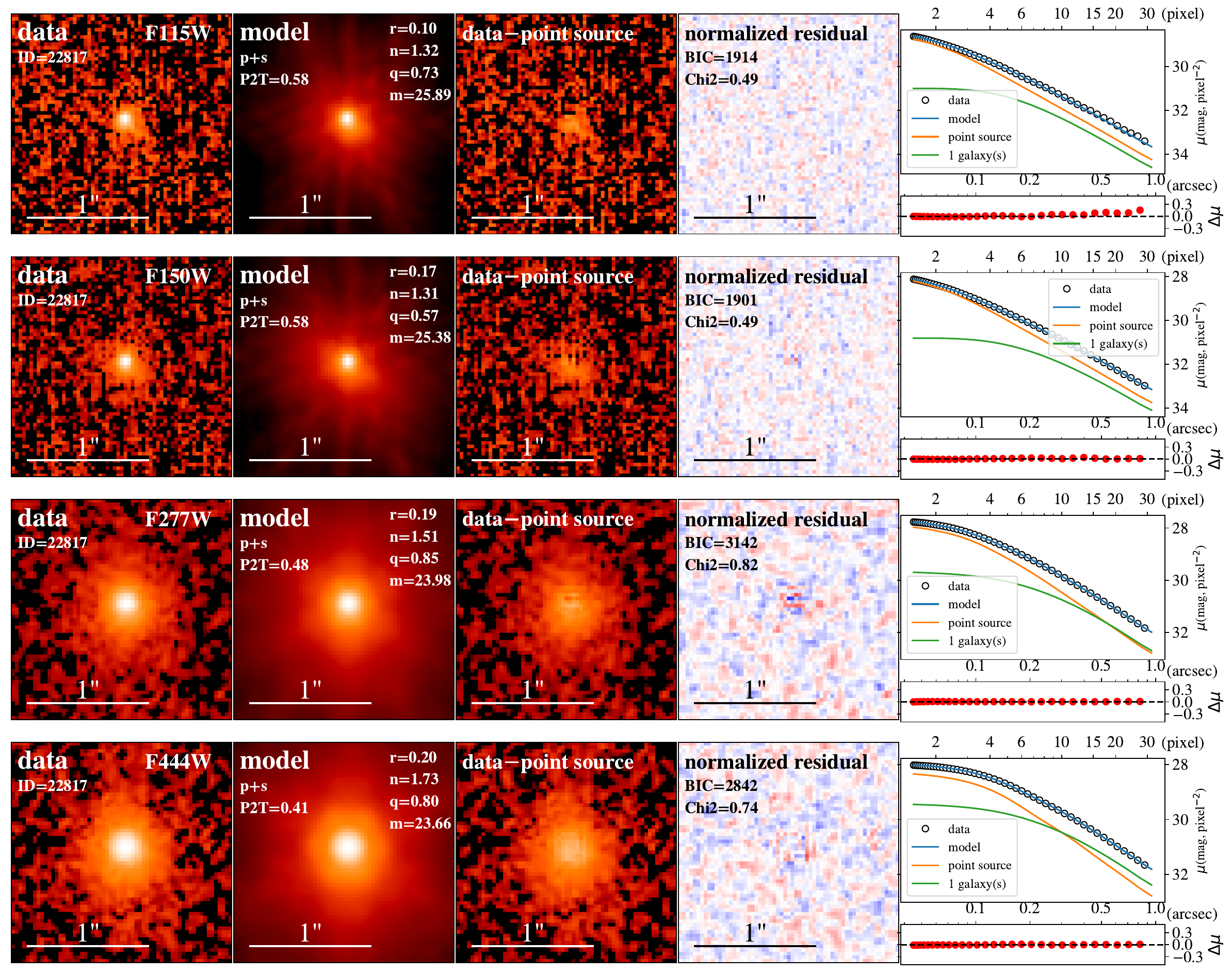}
  \caption{AGN–host decomposition for a representative source (ID 22817) in four JWST/NIRCam bands.\footnote{\url{https://cosmos2025.iap.fr/fitsmap/data/morph_plots/ps1sersic/22817_ps1sersic.pdf}} From top to bottom, the rows correspond to F115W, F150W, F277W, and F444W, respectively. In each row, the panels show, from left to right: (1) observed image; (2) the best-fit two-component model (compact point source + S\'ersic host); (3) the recovered host image after subtraction of the point-source component; (4) the normalized residual map (displayed over $\pm6\sigma$); (5) radial surface brightness profile with data (open circles), total model (blue), point source (orange), and host galaxy (green); the lower panel shows the profile residuals ($\Delta\mu$). Here, $q=b/a$ denotes the projected minor-to-major axis ratio of the S\'ersic host component.
  Decomposition plots for other sources in the COSMOS-Web field are available on the COSMOS-Web data release website.\footnote{\url{https://cosmos2025.iap.fr}} 
  }
  \label{22817_ps1sersic}
\end{figure*}

\section{Method}
\label{method}

\subsection{AGN Sample Construction}
\label{AGN sample construction}

We select AGNs from the AMMIT catalog that have counterparts in COSMOS2025. The cross-matching was performed using sky coordinates with a maximum matching radius of $1''$, retaining the nearest counterpart when multiple matches were found (T. Kelly et al., in preparation). This yields 3,972 AGNs with COSMOS2025 counterparts. For each source, we adopt the best available redshift provided by AMMIT, with spectroscopic redshifts prioritized over photometric estimates. The COSMOS Spectroscopic Redshift Compilation has the highest priority because it incorporates the best available measurements from more than 100 spectroscopic surveys \citep{2026ApJS..282....6K}. Additional spectroscopic redshifts are drawn from DESI-COSMOS, DESI DR1, COSMOS-Legacy, and the VLA-COSMOS 3 GHz Large Project. When no spectroscopic redshift is available, photometric redshifts are adopted preferentially from COSMOS2025, followed by COSMOS2020, COSMOS-Legacy, and VLA-COSMOS. Restricting the sample to $0<z<5$ yields 3,955 AGNs.

Figure~\ref{selection_upset} summarizes the sizes and overlaps of the five AGN selection samples. Of these, 2,654 (67.1\%) are selected by IRAC, 809 (20.5\%) by X-ray, 627 (15.9\%) by radio excess, 341 (8.6\%) by broad emission lines, and 207 (5.2\%) by DESI. Because these samples are not mutually exclusive, their summed counts exceed the total number of unique hosts. Overall, 433 AGNs (10.9\%) are identified by at least two methods, including 233, 152, 46, and 2 identified by exactly two, three, four, and five methods, respectively. Figure~\ref{redshift_distribution} shows the redshift distributions of the full AGN sample and the subsamples identified by different selection methods. The sample is concentrated at $0.5<z<3$, which contains 90.7\% of the plotted sources, with a median redshift of $z=1.78$. IRAC-selected AGNs constitute the largest subsample (67.1\%) and broadly follow the overall redshift distribution, with the selection restricted to $z<3$. In contrast, the X-ray- and radio-selected populations are more concentrated at lower redshifts, with prominent peaks at $z\sim0.9$. The broad-line and DESI subsamples are considerably smaller, with both concentrated primarily at $1\lesssim z\lesssim3$.

\subsection{Stellar Mass Measurement}
\label{mass measurement}

The stellar masses of the AGN host galaxies were estimated using two complementary SED fitting approaches. First, we fitted the multiwavelength photometry with the Code Investigating GALaxy Emission (\texttt{CIGALE}; \citealt{2019A&A...622A.103B}), explicitly including both galaxy and AGN emission components. The stellar masses of the AGN hosts were estimated as likelihood-weighted means over the model distributions. Second, as an independent check, we performed four-band SED fitting using the host-only photometry obtained from the AGN–host image decomposition described in Section~\ref{size measurement}, with Bayesian Analysis of Galaxies for Physical Inference and Parameter EStimation (\texttt{BAGPIPES}; \citealt{2018MNRAS.480.4379C}). 
The latter approach minimizes contamination from unresolved nuclear emission and provides a complementary assessment of the host stellar masses.

For the \texttt{CIGALE} analysis, we adopted the model-based total fluxes and their calibrated uncertainties from the 37-band SourceXtractor++ (SE++; \citealt{2020ASPC..527..461B,2020ASPC..527...29K,2022arXiv221202428K}) photometric catalog of COSMOS2025 \citep{2025A&A...704A.339S}. The photometry spans observed wavelengths of approximately $0.3$--$8\,\mu$m. \texttt{CIGALE} constructs physically motivated SEDs by combining stellar population synthesis, SFHs, nebular emission, dust attenuation and re-emission, and an AGN component. The redshift of each source was fixed during the fitting rather than treated as a free parameter. 
To account for residual systematic uncertainties not captured by the formal photometric errors, we added a 10\% fractional uncertainty in quadrature to the catalog flux density errors. This conservative error floor prevents measurements with very small formal uncertainties from disproportionately influencing the likelihood.

Specifically, the model library was constructed using a delayed SFH, in which the SFR initially rises and subsequently declines exponentially. Stellar emission was modeled using the single stellar population models of \citet{2003MNRAS.344.1000B}, assuming a Chabrier IMF \citep{2003PASP..115..763C}. Nebular line and continuum emission were modeled following \citet{2011MNRAS.415.2920I}, assuming that all ionizing photons are absorbed by the gas. Dust attenuation was assumed to follow the starburst attenuation curve of \citet{2000ApJ...533..682C}, extended to shorter wavelengths following \citet{2002ApJS..140..303L}, with differential attenuation applied to young and old stellar populations. Infrared emission from dust heated by star formation was modeled using the templates of \citet{2014ApJ...784...83D}. The AGN emission was modeled using the SKIRTOR templates \citep{2012MNRAS.420.2756S, 2016MNRAS.458.2288S}, which include emission from the accretion disk, a two-phase dusty torus, and polar dust, allowing for both unobscured and obscured viewing geometries. Finally, intergalactic medium attenuation was applied following \citet{2006MNRAS.365..807M}, and models with ages exceeding that of the Universe at the adopted redshift were excluded.

For each source, we adopted the likelihood-weighted stellar mass, the SFR averaged over the most recent 100 Myr, and the AGN fraction from the CIGALE results. Figure~\ref{42018_best_model} presents the observed photometry and best-fitting SED for a representative source (ID 42018) at $z=2.69$. The model reproduces the observed broadband flux densities over the full wavelength range, with most residuals remaining modest.
The model separates the contributions from stellar emission, nebular emission, dust emission, and AGN emission, with a reduced chi-square of $\chi^2_\nu=0.28$.

To assess the robustness of the \texttt{CIGALE} stellar masses, we performed independent SED fits with \texttt{BAGPIPES} using the host-only photometry obtained from the \texttt{GALIGHT} decomposition. These fits were restricted to the four JWST/NIRCam bands (F115W, F150W, F277W, and F444W). The host magnitudes and their uncertainties were converted into flux densities and corresponding errors \citep{2002MNRAS.329..355A, 2006A&A...457..841I, 2025A&A...704A.339S}.
We adopted the stellar population synthesis models of \citet{2003MNRAS.344.1000B} and the IMF of \citet{2001MNRAS.322..231K}. To prevent measurements with very small formal uncertainties from dominating the likelihood, we imposed a maximum signal-to-noise ratio of 20, corresponding to a minimum fractional flux uncertainty of 5\%. Stellar masses and their credible intervals were derived from the marginalized posterior distributions. 

Given the limited wavelength coverage of the four NIRCam bands, we used the \texttt{BAGPIPES} results primarily to validate the stellar masses derived with \texttt{CIGALE}. The two approaches yield broadly consistent stellar masses, as discussed in Appendix \ref{app:cigale_validation}. We therefore adopted the \texttt{CIGALE} estimates for the subsequent analysis of host-galaxy stellar masses and star formation activity.

\subsection{Size Measurement}
\label{size measurement}

We adopted the structural parameters of AGN host galaxies from the COSMOS-Web morphological catalog described in Section \ref{morphological catalog}. The measurements were obtained using \texttt{GALIGHT}, which decomposes the observed surface brightness into an unresolved point source and an extended S\'ersic component. The point source represents the unresolved nuclear emission, while the S\'ersic component describes the host-galaxy light distribution. Figure \ref{22817_ps1sersic} illustrates the decomposition for a representative source (COSMOS ID 22817).
The decompositions are performed independently in the F115W, F150W, F277W, and F444W bands, providing measurements of the host effective radius ($R_{\rm e}$), S\'ersic index ($n_{\rm S\acute{e}rsic}$), host magnitude, and point-source fraction in each band. The point-source fraction is defined as $f_{\rm PS}=F_{\rm PS}/(F_{\rm PS}+F_{\rm host})$, where $F_{\rm PS}$ and $F_{\rm host}$ are the integrated fluxes of the point source and S\'ersic components, respectively. For the size--mass analysis, we adopt the effective radius and S\'ersic index from the band tracing the rest-frame optical: F115W at $0<z<2$, F150W at $2<z<3$, and F277W at $3<z<5$.

\section{Results} 
\label{results}

\begin{figure*}
  \centering
  \includegraphics[width=\textwidth]{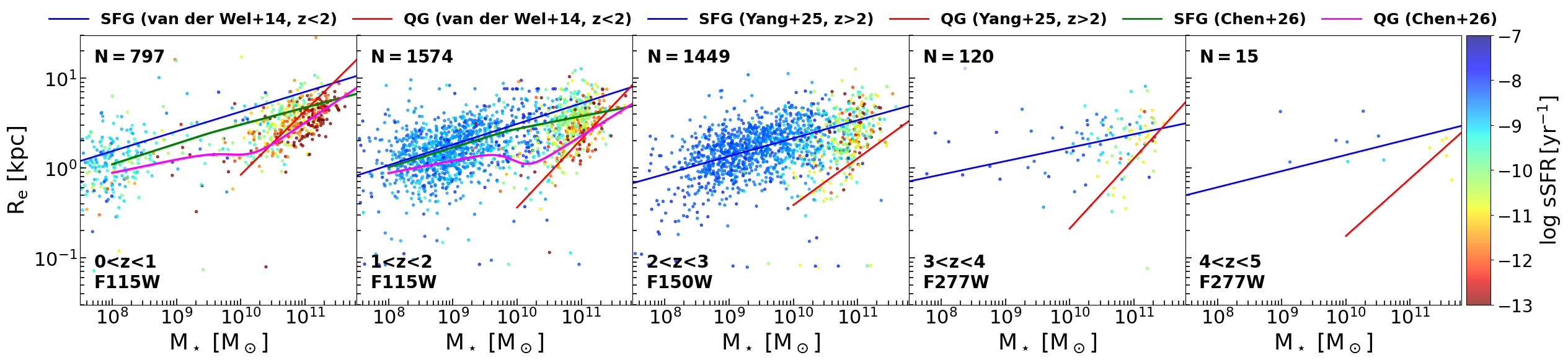}
  \caption{Rest-frame optical size--stellar mass distribution of AGN host galaxies at $0<z<5$ in five redshift bins, color-coded by $\log(\mathrm{sSFR})$. The redshift range and JWST/NIRCam filter used to trace the rest-frame optical emission are indicated in the lower-left corner of each panel, while the number of AGN hosts is shown in the upper-left corner. Blue and red curves represent the size--mass relations of SFGs and QGs, respectively, from \citet{2014ApJ...788...28V} at $0<z<2$ and \citet{2025ApJS..281...68Y} at $2<z<5$. In the two $z<2$ panels, the corresponding SFG and QG relations from \citet{2026ApJ..1002L..40C} are shown in green and magenta, respectively.}
  \label{size_mass_colored_sSFR}
\end{figure*}

\begin{figure*}
  \centering
  \includegraphics[width=\textwidth]{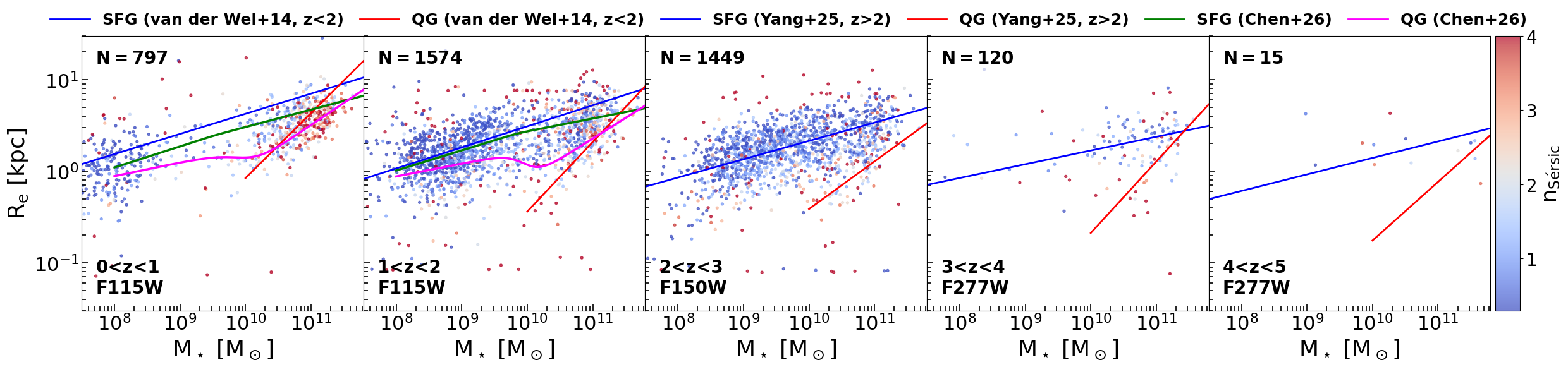}
  \caption{Same as Figure \ref{size_mass_colored_sSFR}, but with points color-coded by S\'ersic index.}
  \label{size_mass_colored_nsersic}
\end{figure*}

\subsection{Size-mass distribution}

We present the rest-frame optical size–mass distributions of AGN host galaxies at $0<z<5$ in Figures~\ref{size_mass_colored_sSFR} and \ref{size_mass_colored_nsersic}, color-coded by sSFR and Sérsic index, respectively. 
For comparison, the best-fit size–mass relations of SFGs and QGs are shown as blue and red lines, respectively, following \citet{2014ApJ...788...28V} at $z<2$ and \citet{2025ApJS..281...68Y} at $z>2$. 
The JWST-based relations at $z<2$ from \citet{2026ApJ..1002L..40C} are also shown as green and magenta lines for SFGs and QGs, respectively.
Across all redshift bins, AGN hosts generally exhibit a positive correlation between effective radius ($R_{\rm e}$) and stellar mass. Their distributions relative to the SFG and QG size--mass relations vary with both redshift and stellar mass.
The $z\sim2$ appears to mark a transition between the higher and lower redshift populations. Within each redshift bin, the behavior of massive hosts with  $\log(M_\star/M_\odot)>10.5$ seems different from lower mass hosts.

At $2<z<5$, the distribution of AGN hosts lies predominantly within the scatter of SFG size–mass relation, but their star formation rate already exhibits a clear dependence on stellar mass. At lower mass regime $\log(M_\star/M_\odot)\leq10.5$, hosts exhibit relatively high sSFRs, i.e., approximately 90.9\% of hosts have $-9<\log(\mathrm{sSFR}/\mathrm{yr}^{-1})<-7$, with a median of $-8.36$. Meanwhile, their S\'ersic indices are also generally low, with a median of $n=0.84$, consistent with disk-dominated morphologies. At higher mass regime, the AGN hosts are still characterized by low S\'ersic indices, with a median of $n=1.27$. However, their star formation activity is already substantially reduced, with a median $\log(\mathrm{sSFR}/\mathrm{yr}^{-1})=-10.25$. These differences between the two mass regimes suggest that at the massive end, the star formation activity of AGN hosts is suppressed whereas their structures are still maintained, i.e., SFG-like size and disk-dominated light profile.

At $1<z<2$, the distinction between lower-mass and massive AGN hosts becomes more pronounced. Most massive hosts lie between the SFG and QG relations, shifting toward the QG relation. Their mean $\log(\mathrm{sSFR}/\mathrm{yr}^{-1})$ is approximately $-10.8$, and their S\'ersic indices span a broad range of $1\lesssim n\lesssim4$. The increasing prevalence of systems with intermediate and high S\'ersic indices indicates a gradual morphological transition from disk-dominated toward bulge-dominated structures. 
In contrast, lower mass hosts remain close to or slightly below the SFG relation, with generally higher sSFRs and lower S\'ersic indices. 
At $0<z<1$, this transition is substantially more advanced. AGN hosts with $\log(M_\star/M_\odot)>10$ lie near or below the QG size-mass relation and exhibit strongly suppressed star formation, with $-13\lesssim\log(\mathrm{sSFR}/\mathrm{yr}^{-1})\lesssim-11$. They are also mainly characterized by high S\'ersic indices, $2\lesssim n\lesssim4$, consistent with bulge-dominated, structurally mature systems. By contrast, lower-mass hosts lie slightly below the SFG relation and generally have $n\lesssim2$, consistent with disk-dominated morphologies. 
It is worth noting that the size offsets should be interpreted with caution. The F115W band probes rest-frame wavelengths redder than the 5000~\AA\ reference wavelength of the \citet{2014ApJ...788...28V} relations, particularly toward the low-redshift end. Wavelength-dependent color gradients may therefore lead to systematically smaller measured effective radii, potentially contributing to the location of some massive AGN hosts below the QG relation \citep{2026arXiv260925886L}.

Taken together, we find the size-mass distribution of AGN hosts relative to the best-fit relation of SFGs and QGs varies across $0<z<5$, mainly caused by the behavior at massive end. The massive AGN hosts with $\log(M_\star/M_\odot)>10.5$, evolve from extended SFG-like systems with reduced star formation at higher redshift ($2<z<3$ and $3<z<4$), through an intermediate regime between SFGs and QGs at $1<z<2$, and eventually toward small sizes, high S\'ersic index and QG-like systems at lower redshift $0<z<1$. 
Their lower-mass counterparts, in contrast, appear to retain star-forming activity and maintain SFGs-like morphology, such as the large size and low S\'ersic index. 

\subsection{Size distribution and its evolution}

\begin{figure*}
  \centering
  \includegraphics[width=\textwidth]{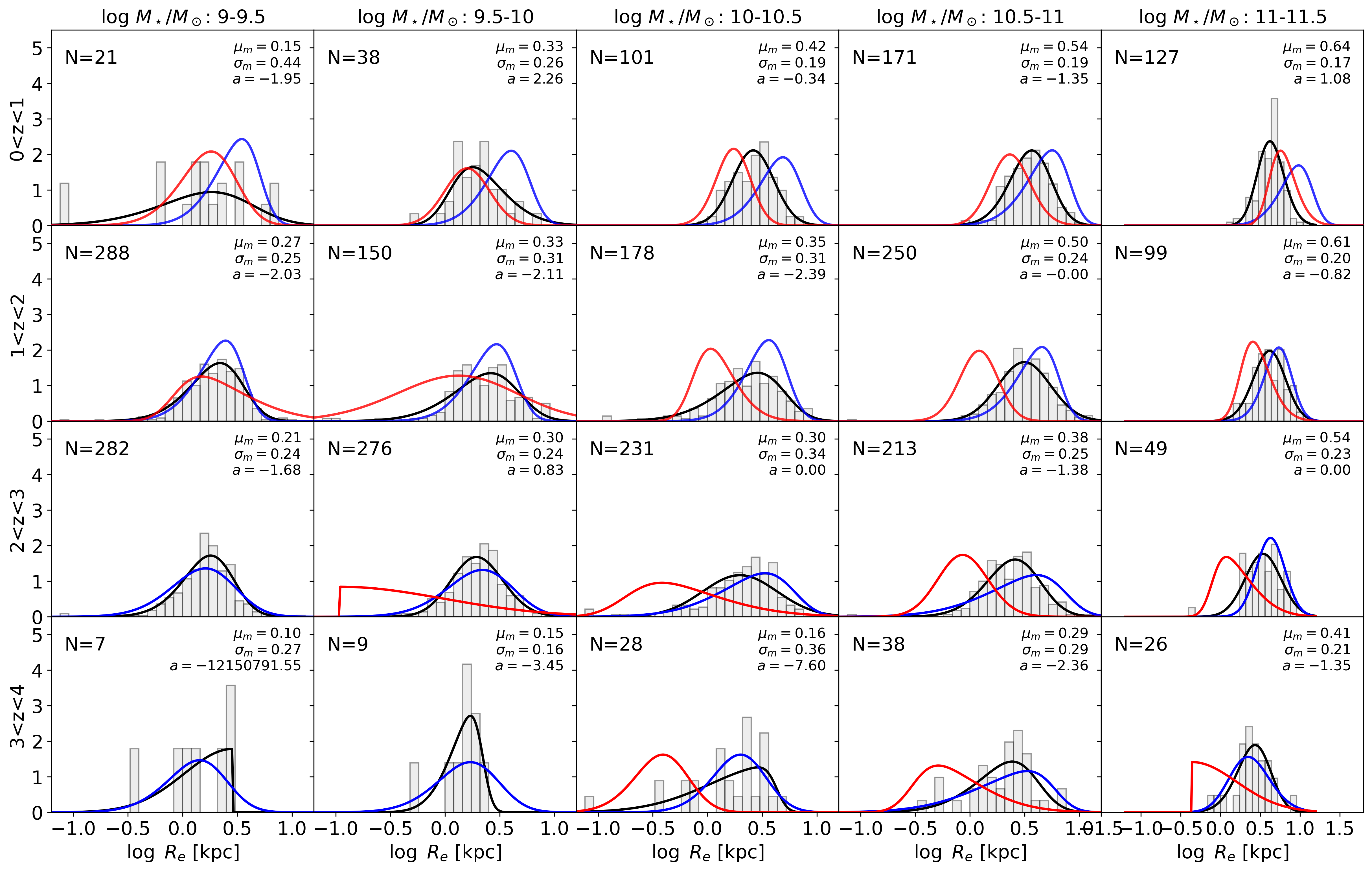}
      \caption{Distributions of the effective radius, $\log (R_{\rm e} /{\rm kpc})$, for AGN host galaxies, shown in bins of redshift (rows) and stellar mass (columns). The light gray histograms show the observed AGN-host distributions, and the solid black curves represent their best fitting skew-normal functions. The number of AGN hosts, $N$, is reported in the upper-left corner of each panel, while the fitted mean $\mu_m$, dispersion $\sigma_m$, and skewness parameter $a$ are given in the upper-right corner. The blue and red curves show the corresponding size distributions expected for SFGs and QGs, respectively, based on the relations of \citet{2014ApJ...788...28V} at $0<z<2$ and \citet{2025ApJS..281...68Y} at $z>2$.}
  \label{Re_hist}
\end{figure*}

\subsubsection{Histograms as a function of stellar mass and redshift}

To further quantify the trends in the size--mass plane, we present the $R_{\rm e}$ distribution of AGN host galaxies in several stellar mass and redshift bins, as shown in Figure~\ref{Re_hist}. The observed distributions are shown as gray histograms and fitted with skewed Gaussian functions using \texttt{scipy.stats.skewnorm}, as shown by the solid black curve. 
The black curves show the best-fitting distributions, and the corresponding mean $\mu_m$, standard deviation $\sigma_m$, and skewness parameter $a$ are reported in each panel. 
For comparison, the blue and red curves show the reference distributions for SFGs and QGs, respectively, from \citet{2014ApJ...788...28V} at $z<2$ and \citet{2025ApJS..281...68Y} at $z>2$. 
Due to the limited sample sizes at $z>4$, we will focus on the results at $0<z<4$ in the following discussion.

In Figure~\ref{Re_hist}, we first find that $\mu_m$ increases with stellar mass within each redshift bin. For example, at $0<z<1$, $\mu_m$ increases from 0.15 to 0.64 across the stellar mass bins, as shown in the top row. Meanwhile, at a fixed stellar mass, $\mu_m$ also tends to increase toward lower redshift. For example, in the $10<\log(M_\star/M_\odot)<10.5$ mass bin (third column), $\mu_m$ increases from 0.16 at $3<z<4$ to 0.42 at $0<z<1$.

We further notice that the location of the AGN-host size distributions relative to those of SFGs and QGs also varies with redshift, particularly at the high-mass end ($\log(M_\star/M_\odot)>10$).
At $2<z<4$, massive AGN hosts have size distributions close to those of SFGs. 
At $1<z<2$, their distributions lie between those of SFGs and QGs, while at $0<z<1$, they overlap more strongly with the QG distributions.
This evolution indicates that the size distribution of massive AGN hosts shifts from an extended, SFG-like population at higher redshifts toward a more compact, QG-like population at later cosmic times, consistent with the trends shown in Figures~\ref{size_mass_colored_sSFR} and \ref{size_mass_colored_nsersic}.
In contrast, the evolution is more gradual at the low-mass end, with $\log(M_\star/M_\odot)<10$. 
Lower mass AGN hosts are predominantly aligned with the SFG distribution at high redshifts. Toward lower redshifts, they increasingly occupy the intermediate region between the SFG and QG distributions but show a less pronounced shift toward the QG population than their massive counterparts.

In addition to the shifts in distribution location, the widths of the size distributions also vary systematically with stellar mass. Within the high-mass regime, $\log(M_\star/M_\odot)>10$, the fitted dispersion generally decreases with increasing stellar mass. At $1<z<4$, $\sigma_m$ decreases from approximately 0.31--0.36 dex at $10<\log(M_\star/M_\odot)<10.5$ to 0.20--0.23 dex at $\log(M_\star/M_\odot)>11$. At $0<z<1$, the high-mass distributions are already relatively narrow, with $\sigma_m\sim0.17$--$0.19$ dex, and the smallest dispersion occurs in the most massive bin. In contrast, the widths of the lower-mass distributions vary more irregularly, without a clear dependence on stellar mass.  
The decreasing width toward higher stellar masses may indicate that the most massive AGN hosts occupy a more restricted range of structural states, whereas lower-mass hosts exhibit greater structural diversity.

\begin{figure}
  \centering
  \includegraphics[width=\columnwidth]{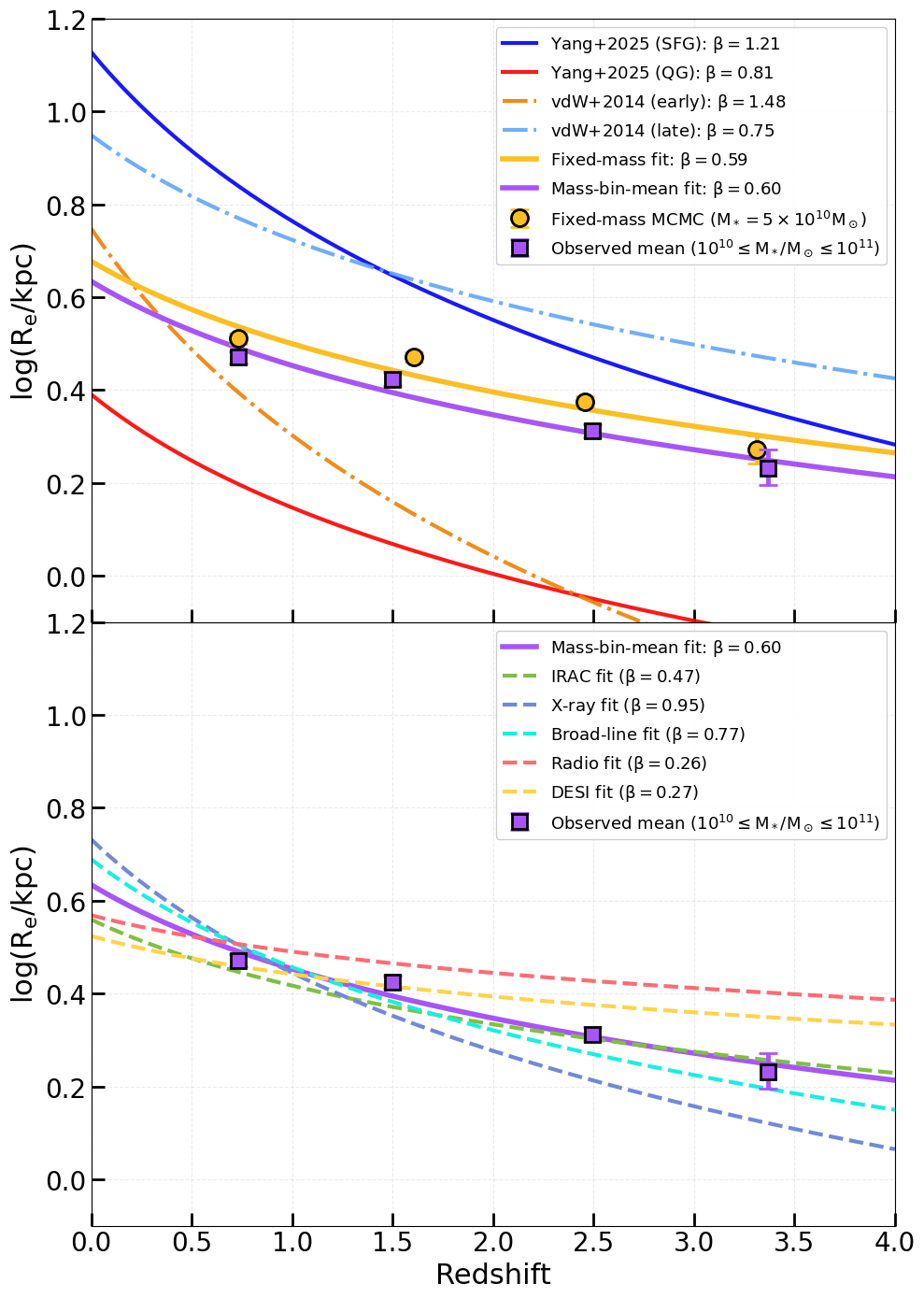}
  \caption{Rest-frame optical size evolution of AGN host galaxies over $0<z<4$. Top: Yellow circles show the posterior-median sizes inferred from MCMC fits to the size–mass relation at $M_\star=5\times10^{10} M_\odot$ in each redshift bin, while purple squares represent the mean sizes of hosts with $10^{10}\leq M_\star/M_\odot\leq10^{11}$. The corresponding solid curves show the best-fitting relations $R_{\rm e}=A(1+z)^{-\beta}$, with $\beta=0.59$ and $0.60$, respectively. Published relations for SFGs and QGs from \citet{2025ApJS..281...68Y}, and for late- and early-type galaxies from \citet{2014ApJ...788...28V}, are shown for comparison. Bottom: The purple squares and solid curve are the same as in the top panel. The colored dashed curves show the size evolution fits for AGNs identified using the five selection methods. 
    }
  \label{size_evolution_different_selections}
\end{figure}

\subsubsection{Mean size evolution}
\label{median size evolution}

We now characterize the mean size evolution of the AGN hosts over $0<z<4$, focusing on systems with $10^{10}\leq M_\star/M_\odot\leq10^{11}$.
The median $\log(M_\star/M_\odot)$ values are 10.65, 10.66, 10.47, and 10.60 in the four redshift bins ($0<z<1$, $1<z<2$, $2<z<3$, and $3<z<4$, respectively), indicating relatively small differences in stellar mass across redshift.
For each redshift bin, we calculate the mean $\log(R_{\rm e}/\mathrm{kpc})$ and place the measurement at the median redshift of the corresponding subsample. 
The uncertainties are estimated from 5,000 bootstrap realizations using the 16th--84th percentile intervals.
We fit the four measurements with $R_{\rm e}=A(1+z)^{-\beta}$ by applying a linear regression to $\log(R_{\rm e}/{\rm kpc})$ as a function of $\log(1+z)$ \citep{2014ApJ...788...28V, 2025ApJS..281...68Y}. The fit yields $\beta_{\rm mean}=0.60\pm0.091$ and $A=4.31\,{\rm kpc}$.

The purple squares and solid curve in the top panel of Figure~\ref{size_evolution_different_selections} show the mean measurements and their best-fitting relation. The mean $\log(R_{\rm e}/\mathrm{kpc})$ decreases from $0.47$ at $z_{\rm med}=0.73$ to $0.23$ at $z_{\rm med}=3.37$. This evolution is shallower than the COSMOS-Web relations for SFGs ($\beta=1.21\pm0.05$) and QGs ($\beta=0.81\pm0.26$) reported by \citet{2025ApJS..281...68Y}, as well as the early- and late-type galaxy relations from \citet{2014ApJ...788...28V}. The mean sizes of AGN hosts generally lie between the COSMOS-Web star-forming and quiescent relations, being closer to the SFG relation at higher redshifts and approaching the QG relation toward lower redshifts.

As a complementary approach, we also examine the evolution of the characteristic size inferred from the size-mass relation. 
Following \citet{2014ApJ...788...28V}, the size-mass relation is parameterized as $\log\left(\frac{R_{\rm e}}{\mathrm{kpc}}\right) =\alpha\log\left(\frac{M_\star}{5\times10^{10}\,M_\odot}\right)+b$, where $\alpha$ is the slope and $b$ represents the characteristic size at the pivot mass $M_\star=5\times10^{10}M_\odot$.
Fitting the posterior estimates of $b$ with $R_{\rm e}=A(1+z)^{-\beta}$ gives $\beta\simeq0.59$. The corresponding measurements and best-fitting relation are shown by the yellow circles and solid curve in the top panel of Figure~\ref{size_evolution_different_selections}. The two methods yield consistent size evolution slopes, with $\Delta\beta\simeq0.01$, supporting a relatively shallow redshift dependence of AGN host sizes.

The relatively shallow size evolution may reflect the changing composition of the AGN-host population, with redshift-dependent contributions from extended, disk-dominated hosts, structurally intermediate systems, and compact, bulge-dominated hosts. At $z\gtrsim2$, the AGN hosts in our sample are predominantly extended and close to the SFG size–mass relation, even when their star formation is already suppressed. Toward lower redshift, the population increasingly overlaps with the compact QG locus. A redshift-dependent mixture of these structural populations can therefore produce a mean size evolution shallower than that of either a uniformly selected SFG or QG population.

The bottom panel of Figure~\ref{size_evolution_different_selections} examines the dependence of the inferred size evolution on AGN selection method. Although all five subsamples exhibit increasing mean size toward lower redshift, their best-fitting slopes span a broad range, from $\beta=0.26$ for radio-selected AGNs to $\beta=0.95$ for X-ray-selected AGNs. The broad-line subsample also shows relatively steep apparent evolution ($\beta=0.77$), whereas the IRAC-selected subsample yields a more moderate slope ($\beta=0.47$), slightly shallower than that of the full sample ($\beta=0.60$). The DESI-selected subsample exhibits similarly weak evolution to the radio-selected subsample, with $\beta=0.27$. Thus, the inferred rate of size evolution depends appreciably on the AGN selection method. This variation may reflect differences in the stellar mass and star formation properties of the selected host populations, as well as their sample sizes and redshift coverage.

\begin{figure*}
  \centering
  \includegraphics[width=\textwidth]{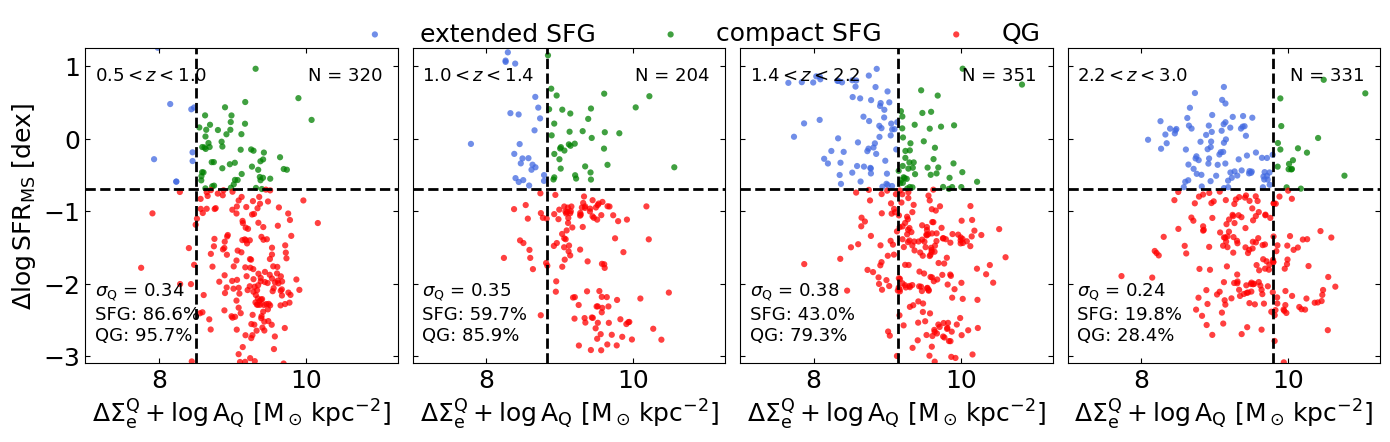}
      \caption{Distribution of $\Delta\log \mathrm{SFR}_{\mathrm{MS}}$ versus effective stellar-mass surface density relative to the quiescent relation for massive AGN host galaxies with $\log(M_\star/M_\odot)>10.3$ across four redshift bins. 
      The x-axis is  $\Delta\Sigma_e^Q+\log A_Q$, where $\log A_Q$ is the zero point of the quiescent $\Sigma_e-M_\star$ relation in each redshift bin. Blue, green, and red points denote extended SFGs, compact SFGs, and QGs, respectively. The horizontal dashed line separates SFGs from QGs at $\Delta\log{\rm SFR}_{\rm MS}=-0.7~\mathrm{dex}$. The vertical dashed line marks the compactness threshold, defined as $-2\sigma_{\rm Q}$ within the QG $\Sigma_e$--$M_\star$ relation, above which galaxies are classified as compact. The values in the upper-right corners give the sample sizes, while the lower-left values give the $1\sigma$ scatter, $\sigma_Q$, of the QG $\Sigma_e-M_\star$ relation and the fractions of SFGs and QGs lying above the compactness threshold. For direct comparison with the canonical L-shaped distribution of massive galaxies, we refer the reader to Figure 6 of \citet{2017ApJ...840...47B}.
    }
  \label{delta_sigmaE_delta_sfr}
\end{figure*}

\section{Discussion}
\label{discussion}

\subsection{AGN Hosts in the SFR–Stellar mass surface density Plane}
\label{AGN Hosts in the SFR–Stellar mass surface density Plane}

The structural evolution of massive galaxies is closely linked to the suppression of star formation during their transition from extended SFGs to compact SFGs.
In the compaction scenario, extended SFGs undergo dissipative gas inflows that drive the formation of compact star-forming systems, known as ``blue nuggets", which may subsequently quench and evolve into compact quiescent systems, or ``red nuggets" \citep[e.g.,][]{2015MNRAS.450.2327Z, 2017ApJ...840...47B}. 
Gas inflows associated with this structural evolution may also fuel AGN activity, suggesting a possible connection between galaxy compaction, black hole growth, and quenching
\citep[e.g.,][]{2017ApJ...846..112K}.
An observational pattern consistent with this evolutionary picture is the approximately L-shaped distribution of SFGs and QGs in the $\Delta\log{\rm SFR}_{\rm MS}$--$\Delta\Sigma_e$ plane as presented in \citet{2017ApJ...840...47B}. 
The $\Sigma_{\rm e}=0.5M_\ast/(\pi R_{\rm e}^{2})$ is effective stellar mass surface density, and $\Delta\Sigma_e$ measures the offset in effective stellar mass surface density from the QG $\Sigma_e$--$M_\star$ relation, while $\Delta\log{\rm SFR}_{\rm MS}$ measures the offset in SFR from the star-forming main sequence. 
This diagram provides a useful framework for examining the connection between structural compactness and star formation quenching. Building on the size--mass relations of AGN hosts presented in Section~\ref{results}, we therefore place our AGN hosts in this plane to investigate how AGN activity is associated with structural compactness and suppressed star formation. Notably, for sources at $0.5<z<3.0$, we calculate $\Sigma_{\rm e}$ using the F150W effective radius of the S\'ersic host component obtained from the image decomposition with \texttt{GALIGHT}. F150W is the closest NIRCam counterpart to the HST/WFC3 F160W band used by \citet{2017ApJ...840...47B}, facilitating a direct comparison with their structural relations.

Figure~\ref{delta_sigmaE_delta_sfr} shows the distribution of massive AGN hosts ($\log(M_\star/M_\odot)>10.3$) in $\Delta\log{\rm SFR}_{\rm MS}$--$\Delta\Sigma_e$ plane across four redshift bins: $0.5<z<1.0$, $1.0<z<1.4$, $1.4<z<2.2$, and $2.2<z<3.0$. 
Specifically, the y-axis in the plot is $\Delta\log {\rm SFR}_{\rm MS}$, while the x-axis is $\Delta\Sigma_{\rm e}^{Q}+\log A_Q$. 
Following \citet{2017ApJ...840...47B}, we also add the QG normalization, $\log A_Q$, taken from their Table~1, so that the redshift evolution of the QG $\Sigma_{\rm e}-M_\star$ relation remains visible. 
We adopt the same thresholds to classify galaxies based on their star-forming activity and compactness.
The AGN hosts with $\Delta\log {\rm SFR}_{\rm MS}>-0.7$ are classified as SFGs, whereas those with $\Delta\log {\rm SFR}_{\rm MS}\leq-0.7$ as QGs. 
As shown in Figure~\ref{delta_sigmaE_delta_sfr}, the horizontal dashed line marks the division between SFGs and QGs at $\Delta\log {\rm SFR}_{\rm MS}=-0.7$.
The vertical dashed line indicates the compactness threshold, $\Delta\Sigma_{\rm e}^{\rm Q}=-2\sigma_{\rm Q}$, relative to the QG $\Sigma_{\rm e}$--$M_\star$ relation; SFGs above/below this threshold are classified as compact/extended SFGs. 
We denote extended SFGs, compact SFGs, and QGs in blue, green, and red, respectively. The lower-left values in each panel give $\sigma_{\rm Q}$ and the fractions of SFGs and QGs lying above the compactness threshold, calculated relative to their respective total populations.

At $2.2<z<3.0$, only 19.8\% of the SFGs lie above the compactness threshold. Although 68.0\% of the AGN hosts in this redshift bin are classified as QGs based on their SFRs, only 28.4\% of these QGs satisfy the compactness criterion.
Thus, a substantial fraction of the high-redshift AGN hosts already exhibit suppressed star formation but still have lower $\Sigma_{\rm e}$ than the typical QG population at the same stellar mass and redshift. 
This suggests that the suppression of star formation and the increase in stellar mass surface density do not always occur together. 
Toward lower redshift, the structural overlap with
QGs increases substantially: from $2.2<z<3.0$ to $0.5<z<1.0$, the fractions above the compactness threshold rise from 28.4\% to 95.7\% for SFR-selected QGs and from 19.8\% to 86.6\% for SFGs.

In the compaction scenario, AGN activity is expected to be enhanced in compact star-forming systems and potentially persist into the quenching phase. 
At $0.5<z<2.2$, most AGN hosts occupy the compact star-forming or compact quiescent regions of the $\Delta\log{\rm SFR}_{\rm MS}$--$\Delta\Sigma_e$ plane, broadly consistent with the compaction scenario.
However, at the high redshift $z>2.2$, we find that our AGN hosts span a broad range in the $\Delta\log{\rm SFR}_{\rm MS}$--$\Delta\Sigma_e$ plane. 
Many hosts with low SFRs still have extended structure, i.e., low $\Sigma_e$. 
This suggests that suppressed star formation in AGN hosts does not 
necessarily coincide with significant structural compactness.
We also acknowledge that the $\Sigma_{\rm e}$ does not uniquely trace central core growth. 
We therefore estimate the stellar-mass surface density within the central 1 kpc, $\Sigma_1$, in Appendix~\ref{app:sigma1}, and we find consistent results.

The absence of a clear L-shaped distribution, particularly the presence of extended AGN hosts with suppressed star formation at high redshift, suggests that the relative timing of star-formation suppression and structural transformation may vary
among AGN hosts. One possible interpretation is that some AGN hosts undergo a different evolutionary pathway, in which star formation is suppressed before substantial structural compaction takes place. 
Future spatially resolved observations of the central gas and stellar mass distributions of AGN hosts will be crucial for clarifying the relationship between AGN activity, star formation quenching, and structural compaction.

\subsection{AGN Selection Methods and Host Galaxy Properties}
\label{selection effects on agn host properties}

\begin{figure*}
  \centering
  \includegraphics[width=\textwidth]{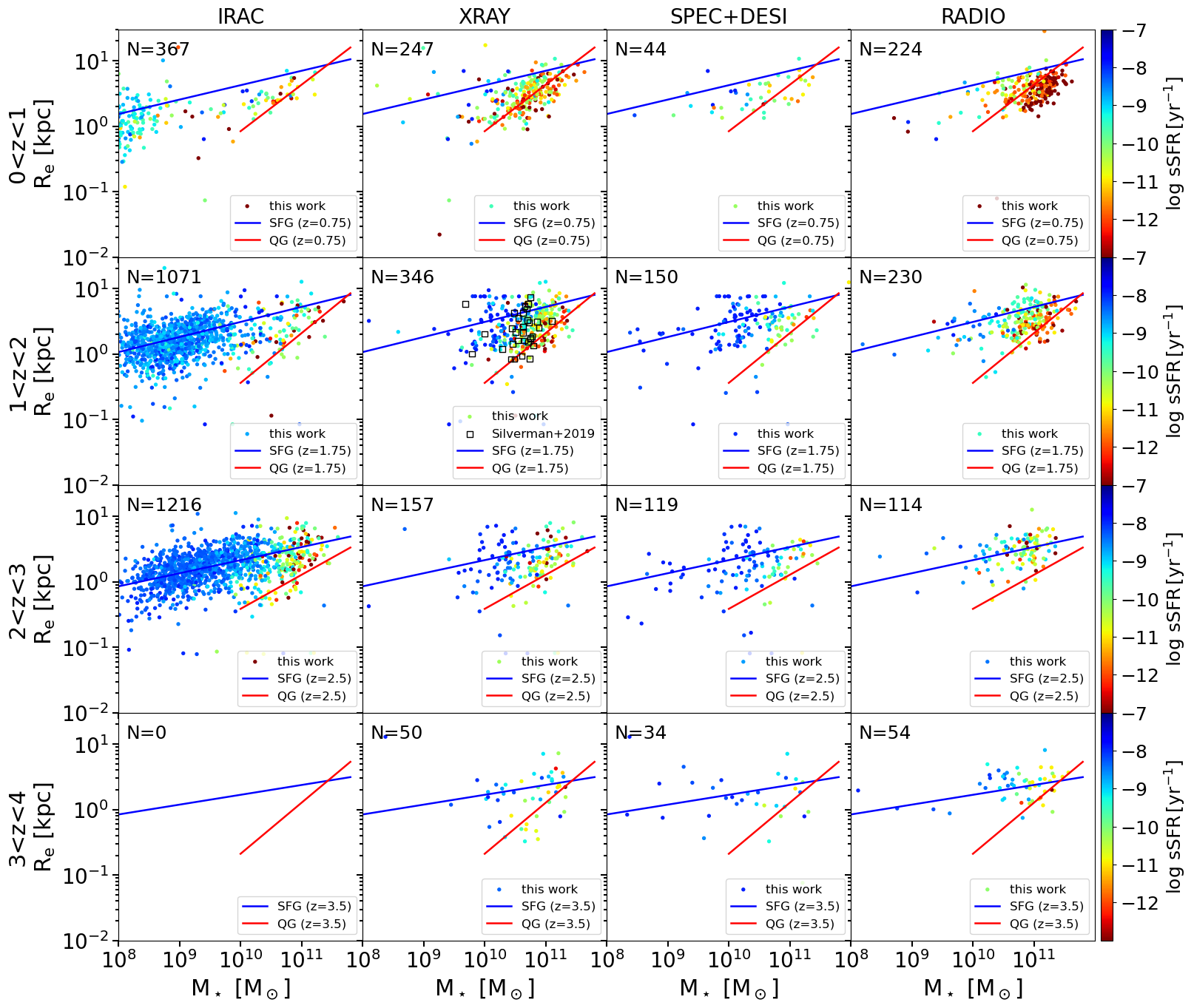}
  \caption{Size–stellar mass distributions of AGN host galaxies in different redshift bins (rows) and for different AGN selection methods (columns), color-coded by sSFR. The columns correspond to AGNs selected by IRAC, X-ray, spectroscopy (SPEC+DESI), and radio, respectively. Blue and red curves show the best-fit relations of SFGs and QGs, respectively. The number of sources in each panel is given in the upper left. Open black squares in the X-ray panel at $1<z<2$ show the HST measurements of 32 X-ray-selected broad-line AGN hosts at $1.2<z<1.7$ from \citet{2019ApJ...887L...5S}.
  }  \label{size_mass_colored_sSFR_selection_method}
\end{figure*}

Previous studies have shown that different AGN selection methods identify partly distinct AGN populations, with radio, X-ray, and infrared selected AGNs showing differences in  host color, morphology, stellar mass, and star-formation activity \citep{2009ApJ...696..891H,2010AJ....140..533G,2018ApJ...853...63D, 2022ApJ...925...74J}. Therefore, no single diagnostic provides a complete census of the AGN population. Studies relying on only one or two selection methods may be biased in their characterization of AGN host galaxies.

We therefore compare the AGN host size--mass relations for the different selection methods in Figure \ref{size_mass_colored_sSFR_selection_method}. Each column represents a different selection method, while the rows correspond to different redshift bins. The subsamples occupy different regions of the size--mass plane and also have different sSFRs. For example, the IRAC-selected AGNs tend to reside in lower-mass hosts with higher sSFRs, whereas the X-ray- and radio-selected samples have higher stellar masses. These differences suggest that each diagnostic probes a different subset of the AGN host population. Below, we discuss the size--mass relation associated with each selection method and consider how the corresponding selection effects may contribute to the observed structural differences and their implications for AGN–host coevolution \citep{2018ARA&A..56..625H}.

Observed-frame IRAC colors select luminous AGNs through the red, approximately power-law mid-infrared continuum, which arises from thermal emission by warm–hot dust in the torus \citep{2012ApJ...748..142D,2018ARA&A..56..625H}. Because mid-infrared emission is relatively insensitive to dust attenuation, IRAC selection can identify obscured AGNs that may be missed by optical selection methods. 
As shown in Figure \ref{size_mass_colored_sSFR_selection_method}, IRAC-selected AGNs are numerically dominated by lower mass hosts, with $\log(M_\star/M_\odot)\lesssim10.5$.
At $2<z<3$, these hosts lie predominantly on or close to the SFG size--mass relation and typically have $\log({\rm sSFR}/{\rm yr}^{-1})\sim-9$ to $-8$. 
At $z<1$, their locus shifts below the SFG relation, indicating smaller effective radii at fixed stellar mass. In contrast, the more massive IRAC-selected hosts exhibit lower sSFRs already at $z>2$, with their distribution spanning the SFG relation and the transitional region between the SFG and QG relations. Thus, IRAC-selected AGN activity in our sample is commonly associated with actively star-forming hosts, particularly at lower stellar masses.  

Compared with IRAC-selected AGNs, the X-ray-selected population is concentrated at higher stellar masses, typically with $\log(M_\star/M_\odot)>10$.
At $2<z<4$, their hosts span the SFG locus and the intermediate region between the SFG and QG size--mass relations, exhibiting a general decline in sSFR. At $1<z<2$, they continue to occupy a broad intermediate region. This distribution is consistent with the HST-based measurements of 32 X-ray-selected broad-line AGN hosts at $1.2<z<1.7$ from \citet{2019ApJ...887L...5S}, shown as open black squares in Figure~\ref{size_mass_colored_sSFR_selection_method}. At $0<z<1$, a subset of the X-ray-selected hosts extends toward the QG relation and exhibits substantially lower sSFRs.

Similarly, the radio-selected AGNs also tend to reside in massive galaxies. Radio emission is dominated by synchrotron radiation associated with the central accretion system and extended jets. Its low optical depth makes radio selection effective for detecting even strongly obscured nuclei \citep{2018ARA&A..56..625H}. 
At $z<1$, radio-selected hosts lie closer to the QG size--mass relation and generally exhibit lower sSFRs than X-ray-selected hosts. At higher redshift, their distribution extends into the intermediate region between the SFG and QG relations and shifts toward the SFG locus, accompanied by higher sSFRs.
This redshift dependence reflects the evolutionary trends of radio-selected AGN host galaxies. In particular, the massive, QG-like hosts at low redshift may be preferentially associated with AGNs with substantial jet activity \citep{2017A&A...602A...3D}. Such jets could help maintain quiescence by heating the surrounding gas and suppressing further cooling \citep{2012ARA&A..50..455F}.

Broad-line AGNs are identified spectroscopically through permitted or semi-forbidden UV and optical emission lines with velocity widths of $\gtrsim1000\,\mathrm{km\,s^{-1}}$. These lines originate from dense, high-velocity gas in the broad-line region close to the supermassive black hole. Because circumnuclear dust can obscure the broad-line region along the line of sight, this method preferentially identifies unobscured Type~1 AGNs \citep{2018ARA&A..56..625H,2026arXiv260612511R}. 
DESI-selected AGNs are identified using a suite of spectroscopic criteria, including broad permitted lines, high-excitation narrow lines, non-stellar continuum emission, BPT and WHAN classifications (T. Kelly et al., in preparation; Juneau et al., in preparation).
Both spectroscopic selections require emission features with sufficient S/N and contrast against the observed continuum and may therefore be incomplete for weak-lined, host-dominated AGNs, including some QGs with low emission-line equivalent widths \citep{2017A&ARv..25....2P,2018ARA&A..56..625H}. Consequently, spectroscopically selected AGN hosts form a relatively sparse distribution in Figure \ref{size_mass_colored_sSFR_selection_method}, clustering primarily along the SFG sequence and transitional region, with limited overlap with the QG relation. This distribution is consistent with spectroscopic selection favoring systems with sufficiently strong emission line signatures and being incomplete for weak-lined, host-dominated AGNs in QGs.

\section{Summary}
\label{summary}

We have presented a comprehensive analysis of the rest-frame optical size-stellar mass relation of 3,955 AGN host galaxies at $0<z<5$ in the COSMOS-Web field. The AGNs are identified through five complementary mid-infrared, X-ray, radio-excess, broad-line, and DESI spectroscopic selections. The effective radii and Sérsic indices of AGN host galaxies are derived from point-source plus S\'ersic model decomposition of JWST/NIRCam images using \texttt{GALIGHT}. Stellar masses and SFRs are inferred by fitting the 37-band photometry spanning 0.3–8 $\mu$m with \texttt{CIGALE}, including an explicit AGN component. Fits to the four-band host-only photometry with \texttt{BAGPIPES} provided an independent check on the stellar mass estimates. Our main results are as follows.

AGN host galaxies exhibit a positive size-mass relation with strong dependencies on stellar mass and redshift.  
At $z\gtrsim2$, massive hosts generally retain extended, disk-dominated, SFG-like structures even when their star formation is already suppressed. Toward lower redshift, they shift from locations near the SFG size--mass relation toward the QG relation, accompanied by declining sSFRs and increasing S\'ersic indices. In contrast, lower-mass hosts evolve more gradually and remain closer to the SFG relation, with relatively high sSFRs and low S\'ersic indices over a broader redshift range.

The size distributions of AGN hosts show clear mass- and redshift-dependent evolution. Their typical sizes increase with stellar mass and toward lower redshift, following the general growth of galaxies over cosmic time. Relative to the normal galaxy population, massive AGN hosts evolve from an SFG-like size distribution at high redshift toward increasing overlap with the QG population at lower redshift, whereas lower mass hosts undergo a weaker transition. The size distributions also become narrower with increasing stellar mass, suggesting a more restricted range of structural states among the most massive hosts and greater structural diversity at lower masses.

AGN hosts exhibit substantially shallower mean size evolution than inactive SFGs and QGs. Over $0<z<4$, the mean sizes of AGN hosts with $10^{10}\leq M_\star/M_\odot\leq10^{11}$ evolve as $R_{\rm e}\propto(1+z)^{-\beta}$, with $\beta\simeq0.60$. An independent estimate based on the characteristic size at $M_\star=5\times10^{10}\,M_\odot$ inferred from the fitted size–mass relations gives $\beta\simeq0.59$. These two estimates agree closely and are both substantially shallower than the corresponding COSMOS-Web trends for inactive SFGs and QGs. The characteristic sizes of AGN hosts generally lie between the COSMOS-Web star-forming and quiescent relations. They are closest to the SFG relation at high redshift but approach the QG and early-type relations toward lower redshift. Among the five AGN selection methods, the inferred size evolution also varies with AGN selection method, being steepest for X-ray-selected AGNs, followed by broad-line-, IRAC-, DESI-, and radio-selected AGNs.

The relative timing of star-formation suppression and structural compaction may vary among AGN hosts. At $z>2.2$, many massive hosts show suppressed star formation, yet the majority remain below the QG compactness threshold. The fraction of AGN hosts above the compactness threshold increases strongly toward lower redshift, demonstrating progressively greater structural overlap with the QG population. The absence in our sample of the clear L-shaped distribution reported for massive galaxies by \citet{2017ApJ...840...47B} suggests that some AGN hosts may follow an alternative evolutionary pathway in which star formation suppression precedes substantial structural compaction.  

The AGN hosts selected via different methods exhibit different properties. IRAC selection primarily identifies host galaxies with lower stellar masses and higher star formation activity, whereas X-ray and radio selections preferentially occupy more massive systems and extend further toward the transitional and quiescent regimes; spectroscopic selections preferentially identify AGNs in galaxies with strong emission lines that are either star-forming or in a transitional state. These systematic differences demonstrate that no single AGN diagnostic provides a complete census of the AGN host population and emphasize the importance of multi-wavelength selection.

\begin{acknowledgments}

We thank the COSMOS-Web collaboration for making the imaging and value-added catalogs used in this work available. This work is based on observations made with the NASA/ESA/CSA James Webb Space Telescope under program GO 1727. The data were obtained from the Mikulski Archive for Space Telescopes at the Space Telescope Science Institute, which is operated by AURA under NASA contract NAS 5-03127. This work was supported by the National Natural Science Foundation of China (Grant No. 12573017). X.D. acknowledges support from the National Key R\&D Program of China (Grant No. 2024YFC2207400). L.C.H. acknowledges support from the National Natural Science Foundation of China (Grant No. 12233001) and the China Manned Space Program (Grant No. CMS-CSST-2025-A09). We also acknowledge the use of COSMOS2025 and other multiwavelength data products and spectroscopic compilations in the COSMOS field, as well as the software GALIGHT \citep{2020ApJ...888...37D}, CIGALE \citep{2019A&A...622A.103B}, BAGPIPES \citep{2018MNRAS.480.4379C}, TOPCAT \citep{2005ASPC..347...29T} and SciPy \citep{2020NaMet..17..261V}.
\end{acknowledgments}

\appendix

\section{Validation of the \texttt{CIGALE} SED Fitting}
\label{app:cigale_validation}

\begin{figure}
  \centering
  \includegraphics[width=\columnwidth]{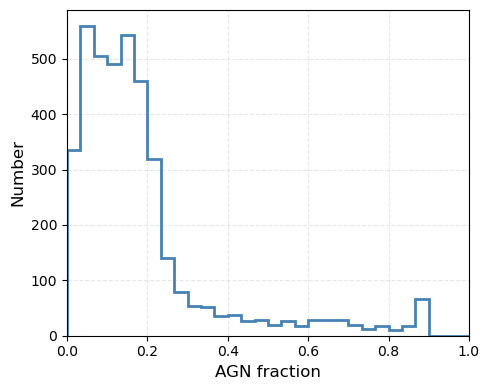}
  \caption{
    Distribution of the \texttt{CIGALE}-derived AGN fraction, defined as the fractional AGN contribution to the total infrared luminosity from dust emission.
  }
  \label{AGN_fraction_cigale}
\end{figure}

\begin{figure*}
  \centering
  \includegraphics[width=6in]{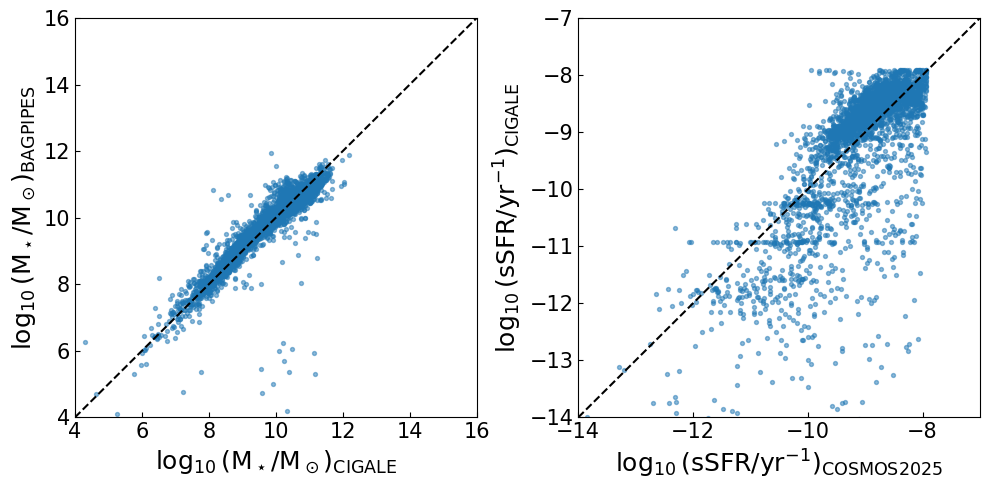}
  \caption{
  Comparison of stellar masses and sSFRs derived from different SED fitting approaches. The left panel compares stellar masses from our \texttt{CIGALE} fits, which include an AGN component, with those from \texttt{BAGPIPES} fits to the host-only photometry after removing the unresolved AGN contribution. The right panel compares our \texttt{CIGALE}-based sSFRs with those reported in COSMOS2025 \citep{2025A&A...704A.339S}. The black dashed lines indicate the one-to-one relation.
  }
  \label{comprison_mass_sSFR_cigale_with_cosmos2025}
\end{figure*}

The SED-fitting framework described in Section~\ref{mass measurement} provides the stellar masses and 100 Myr-averaged SFRs used throughout this work. Here, we examine the inferred nuclear contribution and test the sensitivity of the derived galaxy properties to different SED-modeling assumptions. These analyses are intended to assess the robustness of our measurements rather than to provide an alternative set of physical parameters.

Figure~\ref{AGN_fraction_cigale} presents the distribution of the \texttt{CIGALE}-derived AGN fraction for the full fitted sample. Here, $f_{\rm AGN}$ is defined as the fractional contribution of AGN-heated dust emission to the total dust infrared luminosity. The distribution is strongly skewed toward modest values, with 73\% of sources having $f_{\mathrm{AGN}}\lesssim 0.2$, while a lower-number tail extends to substantially larger AGN contributions. The prevalence of low or moderate values is not inconsistent with the AGN-selected nature of the sample: an AGN may be securely identified through X-ray, infrared, radio, or spectroscopic diagnostics without dominating the total dust infrared luminosity of its host galaxy.

Figure~\ref{comprison_mass_sSFR_cigale_with_cosmos2025} presents two complementary checks of our \texttt{CIGALE} results. The left panel compares our \texttt{CIGALE}-based stellar masses, obtained by modeling both galaxy and AGN emission, with independent \texttt{BAGPIPES} estimates based on host-only photometry after subtracting the unresolved AGN component. The two estimates show strong agreement over a broad mass range, with most sources lying close to the one-to-one relation. This consistency suggests that our stellar mass estimates are relatively insensitive to the treatment of the AGN component and the adopted SED fitting method, supporting their reliability for the size--mass analysis.

The right panel compares our \texttt{CIGALE}-derived sSFRs with the corresponding \texttt{CIGALE} measurements reported in COSMOS2025 \citep{2025A&A...704A.339S}. The comparison reveals appreciably larger scatter, particularly for sources with low inferred sSFRs. This behavior is expected because the recent SFR is more sensitive than stellar mass to the assumed SFH, dust attenuation, and separation of AGN and host emission. The comparison therefore supports the use of the inferred sSFRs for ensemble trends and relative comparisons while cautioning against over-interpreting the precise sSFR of an individual source.

\section{Central Stellar Mass Surface Density within 1 kpc}
\label{app:sigma1}

\begin{figure*}
  \centering
  \includegraphics[width=\textwidth]{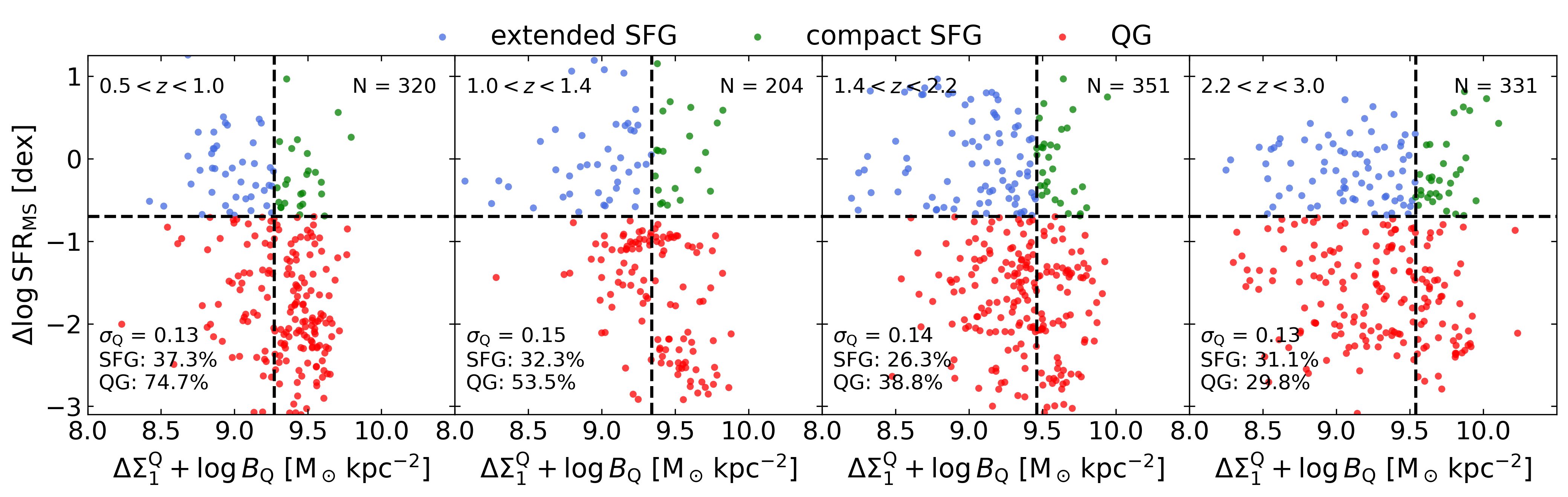}
  \caption{
  Distribution of $\Delta\log{\rm SFR}_{\rm MS}$ versus central stellar-mass surface density relative to the QG structural relation for massive AGN hosts with $\log(M_\star/M_\odot)>10.3$. Blue, green, and red points represent extended SFGs, compact SFGs, and QGs, respectively. The horizontal dashed line separates SFGs and QGs, while the vertical dashed line distinguishes compact from extended galaxies. The upper-right values give the sample sizes, while those in the lower-left indicate the $1\sigma$ scatter ($\sigma_{\rm Q}$) of the QG $\Sigma_1$--$M_\star$ relation and the fractions of SFGs and QGs satisfying the compactness criterion.
  }
  \label{delta_sigma1_delta_sfr}
\end{figure*}

To further examine the central compactness of AGN host galaxies, we calculate the stellar-mass surface density within the central 1 kpc,
\begin{equation}
\Sigma_1 = \frac{M_\star(<1\ {\rm kpc})}{\pi(1\ {\rm kpc})^2}
\end{equation}
We estimate the enclosed stellar mass by integrating the best-fitting F150W S\'ersic host model from \texttt{GALIGHT} within 1 kpc and scaling the resulting enclosed-light fraction by the total stellar mass derived with \texttt{CIGALE}. This procedure assumes a spatially constant stellar mass-to-light ratio. In contrast, \citet{2017ApJ...840...47B} constructed spatially resolved stellar-mass profiles by fitting SEDs to multiband HST surface-brightness profiles and applied an additional size- and S\'ersic-index-dependent correction for PSF smearing within the central 1 kpc. Our estimates, therefore, represent a constant $M_\star/L$ for the S\'ersic host component and do not explicitly account for radial variations in the stellar mass-to-light ratio.

Figure~\ref{delta_sigma1_delta_sfr} shows the distribution of $\Delta\log{\rm SFR}_{\rm MS}$ versus $\Delta\log\Sigma_1^{\rm Q}+\log B_{\rm Q}$ for massive AGN hosts with $\log(M_\star/M_\odot)>10.3$ at $0.5<z<3.0$. Here, $\log B_{\rm Q}$ is the redshift-dependent zero point of the QG $\Sigma_1$--$M_\star$ relation, corresponding to the expected $\log\Sigma_1$ at the pivot mass $M_\star=10^{10.5} M_\odot$ \citep{2017ApJ...840...47B}. At high redshift ($2.2<z<3.0$), the AGN hosts do not exhibit the pronounced L-shaped distribution reported for the general massive galaxy population by \citet{2017ApJ...840...47B}. Only 29.8\% of the SFR-selected QGs lie above the compactness threshold, $\Delta\log\Sigma_1^{\rm Q}=-2\sigma_{\rm Q}$. Thus, many AGN hosts with suppressed star formation have not reached the central stellar-mass densities typical of QGs. This fraction increases to 74.7\% at $0.5<z<1.0$, indicating greater structural overlap with QGs toward lower redshift. By contrast, the corresponding SFG fraction shows little redshift evolution, remaining within $26.3\%$--$37.3\%$.

These trends are broadly consistent with those inferred from $\Sigma_{\rm e}$ in Section~\ref{AGN Hosts in the SFR–Stellar mass surface density Plane}. The absence of a clear L-shaped distribution in either diagnostic may suggest that AGN hosts do not necessarily follow a single evolutionary sequence from extended SFGs through compact SFGs to QGs. Instead, the relative timing of star-formation suppression and structural transformation may vary among AGN hosts, with some systems possibly reaching low SFRs before developing QG-like central densities. This interpretation should remain tentative because our $\Sigma_1$ estimates assume a constant mass-to-light ratio and a S\'ersic host profile and may therefore not fully capture variations in the central stellar-mass distribution.

\bibliography{sample701}

@ARTICLE{2012ApJ...748..142D,
       author = {{Donley}, J.~L. and {Koekemoer}, A.~M. and {Brusa}, M. and {Capak}, P. and {Cardamone}, C.~N. and {Civano}, F. and {Ilbert}, O. and {Impey}, C.~D. and {Kartaltepe}, J.~S. and {Miyaji}, T. and {Salvato}, M. and {Sanders}, D.~B. and {Trump}, J.~R. and {Zamorani}, G.},
        title = "{Identifying Luminous Active Galactic Nuclei in Deep Surveys: Revised IRAC Selection Criteria}",
      journal = {\apj},
         year = 2012,
        month = apr,
       volume = {748},
       number = {2},
          eid = {142},
        pages = {142},
          doi = {10.1088/0004-637X/748/2/142},
archivePrefix = {arXiv},
       eprint = {1201.3899},
 primaryClass = {astro-ph.CO},
       adsurl = {https://ui.adsabs.harvard.edu/abs/2012ApJ...748..142D}
}

@ARTICLE{2023ApJ...954...31C,
       author = {{Casey}, Caitlin M. and {Kartaltepe}, Jeyhan S. and {Drakos}, Nicole E. and {Franco}, Maximilien and {Harish}, Santosh and {Paquereau}, Louise and {Ilbert}, Olivier and {Rose}, Caitlin and {Cox}, Isabella G. and {Nightingale}, James W. and {Robertson}, Brant E. and {Silverman}, John D. and {Koekemoer}, Anton M. and {Massey}, Richard and {McCracken}, Henry Joy and {Rhodes}, Jason and {Akins}, Hollis B. and {Allen}, Natalie and {Amvrosiadis}, Aristeidis and {Arango-Toro}, Rafael C. and {Bagley}, Micaela B. and {Bongiorno}, Angela and {Capak}, Peter L. and {Champagne}, Jaclyn B. and {Chartab}, Nima and {Ch{\'a}vez Ortiz}, {\'O}scar A. and {Chworowsky}, Katherine and {Cooke}, Kevin C. and {Cooper}, Olivia R. and {Darvish}, Behnam and {Ding}, Xuheng and {Faisst}, Andreas L. and {Finkelstein}, Steven L. and {Fujimoto}, Seiji and {Gentile}, Fabrizio and {Gillman}, Steven and {Gould}, Katriona M.~L. and {Gozaliasl}, Ghassem and {Hayward}, Christopher C. and {He}, Qiuhan and {Hemmati}, Shoubaneh and {Hirschmann}, Michaela and {Jahnke}, Knud and {Jin}, Shuowen and {Khostovan}, Ali Ahmad and {Kokorev}, Vasily and {Lambrides}, Erini and {Laigle}, Clotilde and {Larson}, Rebecca L. and {Leung}, Gene C.~K. and {Liu}, Daizhong and {Liaudat}, Tobias and {Long}, Arianna S. and {Magdis}, Georgios and {Mahler}, Guillaume and {Mainieri}, Vincenzo and {Manning}, Sinclaire M. and {Maraston}, Claudia and {Martin}, Crystal L. and {McCleary}, Jacqueline E. and {McKinney}, Jed and {McPartland}, Conor J.~R. and {Mobasher}, Bahram and {Pattnaik}, Rohan and {Renzini}, Alvio and {Rich}, R. Michael and {Sanders}, David B. and {Sattari}, Zahra and {Scognamiglio}, Diana and {Scoville}, Nick and {Sheth}, Kartik and {Shuntov}, Marko and {Sparre}, Martin and {Suzuki}, Tomoko L. and {Talia}, Margherita and {Toft}, Sune and {Trakhtenbrot}, Benny and {Urry}, C. Megan and {Valentino}, Francesco and {Vanderhoof}, Brittany N. and {Vardoulaki}, Eleni and {Weaver}, John R. and {Whitaker}, Katherine E. and {Wilkins}, Stephen M. and {Yang}, Lilan and {Zavala}, Jorge A.},
        title = "{COSMOS-Web: An Overview of the JWST Cosmic Origins Survey}",
      journal = {\apj},
         year = 2023,
        month = sep,
       volume = {954},
       number = {1},
          eid = {31},
        pages = {31},
          doi = {10.3847/1538-4357/acc2bc},
archivePrefix = {arXiv},
       eprint = {2211.07865},
 primaryClass = {astro-ph.GA},
       adsurl = {https://ui.adsabs.harvard.edu/abs/2023ApJ...954...31C}
}

@ARTICLE{2025arXiv250201724D,
       author = {{De Lucia}, Gabriella and {Fontanot}, Fabio and {Hirschmann}, Michaela and {Xie}, Lizhi},
        title = "{Cosmic quenching}",
      journal = {arXiv e-prints},
         year = 2025,
        month = feb,
          eid = {arXiv:2502.01724},
        pages = {arXiv:2502.01724},
          doi = {10.48550/arXiv.2502.01724},
archivePrefix = {arXiv},
       eprint = {2502.01724},
 primaryClass = {astro-ph.GA},
       adsurl = {https://ui.adsabs.harvard.edu/abs/2025arXiv250201724D}
}

@ARTICLE{2026arXiv260612511R,
       author = {{Ricci}, Claudio},
        title = "{Unification models of Active Galactic Nuclei}",
      journal = {arXiv e-prints},
         year = 2026,
        month = jun,
          eid = {arXiv:2606.12511},
        pages = {arXiv:2606.12511},
          doi = {10.48550/arXiv.2606.12511},
archivePrefix = {arXiv},
       eprint = {2606.12511},
 primaryClass = {astro-ph.GA},
       adsurl = {https://ui.adsabs.harvard.edu/abs/2026arXiv260612511R}
}

@ARTICLE{2017A&ARv..25....2P,
       author = {{Padovani}, P. and {Alexander}, D.~M. and {Assef}, R.~J. and {De Marco}, B. and {Giommi}, P. and {Hickox}, R.~C. and {Richards}, G.~T. and {Smol{\v{c}}i{\'c}}, V. and {Hatziminaoglou}, E. and {Mainieri}, V. and {Salvato}, M.},
        title = "{Active galactic nuclei: what's in a name?}",
      journal = {\aapr},
         year = 2017,
        month = aug,
       volume = {25},
       number = {1},
          eid = {2},
        pages = {2},
          doi = {10.1007/s00159-017-0102-9},
archivePrefix = {arXiv},
       eprint = {1707.07134},
 primaryClass = {astro-ph.GA},
       adsurl = {https://ui.adsabs.harvard.edu/abs/2017A&ARv..25....2P}
}

@ARTICLE{1996A&AS..117..393B,
       author = {{Bertin}, E. and {Arnouts}, S.},
        title = "{SExtractor: Software for source extraction.}",
      journal = {\aaps},
         year = 1996,
        month = jun,
       volume = {117},
        pages = {393-404},
          doi = {10.1051/aas:1996164},
       adsurl = {https://ui.adsabs.harvard.edu/abs/1996A&AS..117..393B}
}

@article{Barbary2016, doi = {10.21105/joss.00058}, url = {https://doi.org/10.21105/joss.00058}, year = {2016}, publisher = {The Open Journal}, volume = {1}, number = {6}, pages = {58}, author = {Barbary, Kyle}, title = {SEP: Source Extractor as a library}, journal = {Journal of Open Source Software} }

@ARTICLE{2024ApJ...962...93Z,
       author = {{Zhuang}, Ming-Yang and {Li}, Junyao and {Shen}, Yue},
        title = "{Active Galactic Nuclei and Host Galaxies in COSMOS-Web. I. NIRCam Images, Point-spread-function Models and Initial Results on X-Ray-selected Broad-line AGNs at 0.35 {\ensuremath{\lesssim}} z {\ensuremath{\lesssim}} 3.5}",
      journal = {\apj},
         year = 2024,
        month = feb,
       volume = {962},
       number = {1},
          eid = {93},
        pages = {93},
          doi = {10.3847/1538-4357/ad1517},
archivePrefix = {arXiv},
       eprint = {2309.03266},
 primaryClass = {astro-ph.GA},
       adsurl = {https://ui.adsabs.harvard.edu/abs/2024ApJ...962...93Z}
}

@ARTICLE{2026ApJ...999...70L,
       author = {{Li}, Ruancun and {Ho}, Luis C. and {Chen}, Chang-Hao},
        title = "{The Dichotomy in the Nuclear and Host-galaxy Properties of High-redshift Quasars}",
      journal = {\apj},
         year = 2026,
        month = mar,
       volume = {999},
       number = {1},
          eid = {70},
        pages = {70},
          doi = {10.3847/1538-4357/ae38b3},
archivePrefix = {arXiv},
       eprint = {2505.12867},
 primaryClass = {astro-ph.GA},
       adsurl = {https://ui.adsabs.harvard.edu/abs/2026ApJ...999...70L}
}

@ARTICLE{2014ARA&A..52..291C,
       author = {{Conselice}, Christopher J.},
        title = "{The Evolution of Galaxy Structure Over Cosmic Time}",
      journal = {\araa},
         year = 2014,
        month = aug,
       volume = {52},
        pages = {291-337},
          doi = {10.1146/annurev-astro-081913-040037},
archivePrefix = {arXiv},
       eprint = {1403.2783},
 primaryClass = {astro-ph.GA},
       adsurl = {https://ui.adsabs.harvard.edu/abs/2014ARA&A..52..291C}
}

@ARTICLE{2011ApJ...742...61S,
       author = {{Salvato}, M. and {Ilbert}, O. and {Hasinger}, G. and {Rau}, A. and {Civano}, F. and {Zamorani}, G. and {Brusa}, M. and {Elvis}, M. and {Vignali}, C. and {Aussel}, H. and {Comastri}, A. and {Fiore}, F. and {Le Floc'h}, E. and {Mainieri}, V. and {Bardelli}, S. and {Bolzonella}, M. and {Bongiorno}, A. and {Capak}, P. and {Caputi}, K. and {Cappelluti}, N. and {Carollo}, C.~M. and {Contini}, T. and {Garilli}, B. and {Iovino}, A. and {Fotopoulou}, S. and {Fruscione}, A. and {Gilli}, R. and {Halliday}, C. and {Kneib}, J.-P. and {Kakazu}, Y. and {Kartaltepe}, J.~S. and {Koekemoer}, A.~M. and {Kovac}, K. and {Ideue}, Y. and {Ikeda}, H. and {Impey}, C.~D. and {Le Fevre}, O. and {Lamareille}, F. and {Lanzuisi}, G. and {Le Borgne}, J.-F. and {Le Brun}, V. and {Lilly}, S. and {Maier}, C. and {Manohar}, S. and {Masters}, D. and {McCracken}, H. and {Messias}, H. and {Mignoli}, M. and {Mobasher}, B. and {Nagao}, T. and {Pello}, R. and {Puccetti}, S. and {Perez-Montero}, E. and {Renzini}, A. and {Sargent}, M. and {Sanders}, D.~B. and {Scodeggio}, M. and {Scoville}, N. and {Shopbell}, P. and {Silvermann}, J. and {Taniguchi}, Y. and {Tasca}, L. and {Tresse}, L. and {Trump}, J.~R. and {Zucca}, E.},
        title = "{Dissecting Photometric Redshift for Active Galactic Nucleus Using XMM- and Chandra-COSMOS Samples}",
      journal = {\apj},
         year = 2011,
        month = dec,
       volume = {742},
       number = {2},
          eid = {61},
        pages = {61},
          doi = {10.1088/0004-637X/742/2/61},
archivePrefix = {arXiv},
       eprint = {1108.6061},
 primaryClass = {astro-ph.CO},
       adsurl = {https://ui.adsabs.harvard.edu/abs/2011ApJ...742...61S}
}

@ARTICLE{2022A&A...667A..65K,
       author = {{Kauffmann}, O.~B. and {Ilbert}, O. and {Weaver}, J.~R. and {McCracken}, H.~J. and {Milvang-Jensen}, B. and {Brammer}, G. and {Davidzon}, I. and {Le F{\`e}vre}, O. and {Liu}, D. and {Mobasher}, B. and {Moneti}, A. and {Shuntov}, M. and {Toft}, S. and {Casey}, C.~M. and {Dunlop}, J.~S. and {Kartaltepe}, J.~S. and {Koekemoer}, A.~M. and {Sanders}, D.~B. and {Tresse}, L.},
        title = "{COSMOS2020: UV-selected galaxies at z {\ensuremath{\geq}} 7.5}",
      journal = {\aap},
         year = 2022,
        month = nov,
       volume = {667},
          eid = {A65},
        pages = {A65},
          doi = {10.1051/0004-6361/202243088},
archivePrefix = {arXiv},
       eprint = {2207.11740},
 primaryClass = {astro-ph.GA},
       adsurl = {https://ui.adsabs.harvard.edu/abs/2022A&A...667A..65K}
}

@ARTICLE{2023ApJ...946L..14K,
       author = {{Kocevski}, Dale D. and {Barro}, Guillermo and {McGrath}, Elizabeth J. and {Finkelstein}, Steven L. and {Bagley}, Micaela B. and {Ferguson}, Henry C. and {Jogee}, Shardha and {Yang}, Guang and {Dickinson}, Mark and {Hathi}, Nimish P. and {Backhaus}, Bren E. and {Bell}, Eric F. and {Bisigello}, Laura and {Buat}, V{\'e}ronique and {Burgarella}, Denis and {Casey}, Caitlin M. and {Cleri}, Nikko J. and {Cooper}, M.~C. and {Costantin}, Luca and {Croton}, Darren and {Daddi}, Emanuele and {Fontana}, Adriano and {Fujimoto}, Seiji and {Gardner}, Jonathan P. and {Gawiser}, Eric and {Giavalisco}, Mauro and {Grazian}, Andrea and {Grogin}, Norman A. and {Guo}, Yuchen and {Arrabal Haro}, Pablo and {Hirschmann}, Michaela and {Holwerda}, Benne W. and {Huertas-Company}, Marc and {Hutchison}, Taylor A. and {Iyer}, Kartheik G. and {Jones}, Brenda and {Juneau}, St{\'e}phanie and {Kartaltepe}, Jeyhan S. and {Kewley}, Lisa J. and {Kirkpatrick}, Allison and {Koekemoer}, Anton M. and {Kurczynski}, Peter and {Le Bail}, Aur{\'e}lien and {Long}, Arianna S. and {Lotz}, Jennifer M. and {Lucas}, Ray A. and {Papovich}, Casey and {Pentericci}, Laura and {P{\'e}rez-Gonz{\'a}lez}, Pablo G. and {Pirzkal}, Nor and {Rafelski}, Marc and {Ravindranath}, Swara and {Somerville}, Rachel S. and {Straughn}, Amber N. and {Tacchella}, Sandro and {Trump}, Jonathan R. and {Wilkins}, Stephen M. and {Wuyts}, Stijn and {Yung}, L.~Y. Aaron and {Zavala}, Jorge A.},
        title = "{CEERS Key Paper. II. A First Look at the Resolved Host Properties of AGN at 3 < z < 5 with JWST}",
      journal = {\apjl},
         year = 2023,
        month = mar,
       volume = {946},
       number = {1},
          eid = {L14},
        pages = {L14},
          doi = {10.3847/2041-8213/acad00},
archivePrefix = {arXiv},
       eprint = {2208.14480},
 primaryClass = {astro-ph.GA},
       adsurl = {https://ui.adsabs.harvard.edu/abs/2023ApJ...946L..14K}
}

@ARTICLE{2026arXiv260925886L,
       author = {{Laishram}, Ronaldo and {Kusakabe}, Haruka and {Kikuta}, Satoshi and {Shimizu}, Shunta and {Koyama}, Yusei and {Kodama}, Tadayuki and {Hirschmann}, Michaela and {Toni}, Greta and {Flayhart}, Carter and {Samir}, Rasha M. and {Saputra Haryana}, Novan},
        title = "{Rest-Frame UV and Optical Structural Evolution of Narrowband-Selected Lyman-alpha Emitters at z = 2-7 with JWST/NIRCam}",
      journal = {arXiv e-prints},
         year = 2026,
        month = sep,
          eid = {arXiv:2609.25886},
        pages = {arXiv:2609.25886},
          doi = {10.48550/arXiv.2609.25886},
archivePrefix = {arXiv},
       eprint = {2609.25886},
 primaryClass = {astro-ph.GA},
       adsurl = {https://ui.adsabs.harvard.edu/abs/2026arXiv260925886L}
}

@ARTICLE{1983ApJ...266..713O,
       author = {{Oke}, J.~B. and {Gunn}, J.~E.},
        title = "{Secondary standard stars for absolute spectrophotometry.}",
      journal = {\apj},
         year = 1983,
        month = mar,
       volume = {266},
        pages = {713-717},
          doi = {10.1086/160817},
       adsurl = {https://ui.adsabs.harvard.edu/abs/1983ApJ...266..713O}
}

@ARTICLE{2026ApJ..1009..106H,
       author = {{Haryana}, Novan Saputra and {Akiyama}, Masayuki and {Abdurro'uf} and {Cooray}, Suchetha and {Fitriana}, Itsna Khoirul and {Triani}, Dian P. and {Vijarnwannaluk}, Bovornpratch and {Alfonzo}, Juan Pablo and {Andika}, Irham Taufik and {Effendi}, Muhammad Nur Ihsan and {Huda}, Ibnu Nurul and {Kakimoto}, Takumi and {Jaelani}, Anton Timur and {Laishram}, Ronaldo and {Matsumoto}, Naoki and {Naufal}, Abdurrahman and {Puspitarini}, Lucky and {Sutanto}, Ryo Albert and {Wulandari}, Hesti Retno Tri},
        title = "{Interplay of Compaction, Quenching, and Black Hole Growth in the Most Massive Galaxies since z \raisebox{-0.5ex}\textasciitilde 5: Insights from JWST and Chandra Data}",
      journal = {\apj},
         year = 2026,
        month = oct,
       volume = {1009},
       number = {2},
          eid = {106},
        pages = {106},
          doi = {10.3847/1538-4357/ae9f49},
archivePrefix = {arXiv},
       eprint = {2608.24124},
 primaryClass = {astro-ph.GA},
       adsurl = {https://ui.adsabs.harvard.edu/abs/2026ApJ..1009..106H}
}

@ARTICLE{2015ARA&A..53...51S,
       author = {{Somerville}, Rachel S. and {Dav{\'e}}, Romeel},
        title = "{Physical Models of Galaxy Formation in a Cosmological Framework}",
      journal = {\araa},
         year = 2015,
        month = aug,
       volume = {53},
        pages = {51-113},
          doi = {10.1146/annurev-astro-082812-140951},
archivePrefix = {arXiv},
       eprint = {1412.2712},
 primaryClass = {astro-ph.GA},
       adsurl = {https://ui.adsabs.harvard.edu/abs/2015ARA&A..53...51S}
}

@ARTICLE{2026ApJ...997..366K,
       author = {{Kelly}, Thresa and {Auge}, Connor and {Sanders}, David and {Kirkpatrick}, Allison and {Carroll}, Christopher M. and {Urry}, C. Megan and {Cappelluti}, Nico and {Ananna}, Tonima Tasnim and {Balokovi{\'c}}, Mislav and {Boquien}, M{\'e}d{\'e}ric and {Coleman}, Brandon and {Hickox}, Ryan and {Kartaltepe}, Jeyhan S. and {Koss}, Michael and {Marchesi}, Stefano and {Peca}, Alessandro and {Powell}, Meredith and {Treister}, Ezequiel and {Troiani}, Greg and {Turner}, Tracey Jane and {Weaver}, Tristan},
        title = "{Accretion History of AGNs: A Multiwavelength Comparison of X-Ray- and Infrared-selected AGNs}",
      journal = {\apj},
         year = 2026,
        month = feb,
       volume = {997},
       number = {2},
          eid = {366},
        pages = {366},
          doi = {10.3847/1538-4357/ae2a32},
       adsurl = {https://ui.adsabs.harvard.edu/abs/2026ApJ...997..366K}
}

@ARTICLE{2017ApJ...846..112K,
       author = {{Kocevski}, Dale D. and {Barro}, Guillermo and {Faber}, S.~M. and {Dekel}, Avishai and {Somerville}, Rachel S. and {Young}, Joshua A. and {Williams}, Christina C. and {McIntosh}, Daniel H. and {Georgakakis}, Antonis and {Hasinger}, Guenther and {Nandra}, Kirpal and {Civano}, Francesca and {Alexander}, David M. and {Almaini}, Omar and {Conselice}, Christopher J. and {Donley}, Jennifer L. and {Ferguson}, Harry C. and {Giavalisco}, Mauro and {Grogin}, Norman A. and {Hathi}, Nimish and {Hawkins}, Matthew and {Koekemoer}, Anton M. and {Koo}, David C. and {McGrath}, Elizabeth J. and {Mobasher}, Bahram and {P{\'e}rez Gonz{\'a}lez}, Pablo G. and {Pforr}, Janine and {Primack}, Joel R. and {Santini}, Paola and {Stefanon}, Mauro and {Trump}, Jonathan R. and {van der Wel}, Arjen and {Wuyts}, Stijn and {Yan}, Haojing},
        title = "{CANDELS: Elevated Black Hole Growth in the Progenitors of Compact Quiescent Galaxies at z {\ensuremath{\sim}} 2}",
      journal = {\apj},
         year = 2017,
        month = sep,
       volume = {846},
       number = {2},
          eid = {112},
        pages = {112},
          doi = {10.3847/1538-4357/aa8566},
archivePrefix = {arXiv},
       eprint = {1710.05921},
 primaryClass = {astro-ph.GA},
       adsurl = {https://ui.adsabs.harvard.edu/abs/2017ApJ...846..112K}
}

@ARTICLE{2022MNRAS.515.4860A,
       author = {{Aird}, James and {Coil}, Alison L. and {Kocevski}, Dale D.},
        title = "{AGN accretion and black hole growth across compact and extended galaxy evolution phases}",
      journal = {\mnras},
         year = 2022,
        month = oct,
       volume = {515},
       number = {4},
        pages = {4860-4889},
          doi = {10.1093/mnras/stac2103},
archivePrefix = {arXiv},
       eprint = {2201.11756},
 primaryClass = {astro-ph.GA},
       adsurl = {https://ui.adsabs.harvard.edu/abs/2022MNRAS.515.4860A}
}

@ARTICLE{2017ApJ...840...47B,
       author = {{Barro}, Guillermo and {Faber}, S.~M. and {Koo}, David C. and {Dekel}, Avishai and {Fang}, Jerome J. and {Trump}, Jonathan R. and {P{\'e}rez-Gonz{\'a}lez}, Pablo G. and {Pacifici}, Camilla and {Primack}, Joel R. and {Somerville}, Rachel S. and {Yan}, Haojing and {Guo}, Yicheng and {Liu}, Fengshan and {Ceverino}, Daniel and {Kocevski}, Dale D. and {McGrath}, Elizabeth},
        title = "{Structural and Star-forming Relations since z {\ensuremath{\sim}} 3: Connecting Compact Star-forming and Quiescent Galaxies}",
      journal = {\apj},
         year = 2017,
        month = may,
       volume = {840},
       number = {1},
          eid = {47},
        pages = {47},
          doi = {10.3847/1538-4357/aa6b05},
archivePrefix = {arXiv},
       eprint = {1509.00469},
 primaryClass = {astro-ph.GA},
       adsurl = {https://ui.adsabs.harvard.edu/abs/2017ApJ...840...47B}
}

@ARTICLE{2018ApJ...853...63D,
       author = {{Donley}, J.~L. and {Kartaltepe}, J. and {Kocevski}, D. and {Salvato}, M. and {Santini}, P. and {Suh}, H. and {Civano}, F. and {Koekemoer}, A.~M. and {Trump}, J. and {Brusa}, M. and {Cardamone}, C. and {Castro}, A. and {Cisternas}, M. and {Conselice}, C. and {Croton}, D. and {Hathi}, N. and {Liu}, C. and {Lucas}, R.~A. and {Nair}, P. and {Rosario}, D. and {Sanders}, D. and {Simmons}, B. and {Villforth}, C. and {Alexander}, D.~M. and {Bell}, E.~F. and {Faber}, S.~M. and {Grogin}, N.~A. and {Lotz}, J. and {McIntosh}, D.~H. and {Nagao}, T.},
        title = "{Evidence for Merger-driven Growth in Luminous, High-z, Obscured AGNs in the CANDELS/COSMOS Field}",
      journal = {\apj},
         year = 2018,
        month = jan,
       volume = {853},
       number = {1},
          eid = {63},
        pages = {63},
          doi = {10.3847/1538-4357/aa9ffa},
archivePrefix = {arXiv},
       eprint = {1712.02424},
 primaryClass = {astro-ph.GA},
       adsurl = {https://ui.adsabs.harvard.edu/abs/2018ApJ...853...63D}
}

@ARTICLE{2022ApJ...925...74J,
       author = {{Ji}, Zhiyuan and {Giavalisco}, Mauro and {Kirkpatrick}, Allison and {Kocevski}, Dale and {Daddi}, Emanuele and {Delvecchio}, Ivan and {Hatcher}, Cassandra},
        title = "{AGN Selection Methods Have Profound Impacts on the Distributions of Host-galaxy Properties}",
      journal = {\apj},
         year = 2022,
        month = jan,
       volume = {925},
       number = {1},
          eid = {74},
        pages = {74},
          doi = {10.3847/1538-4357/ac3837},
archivePrefix = {arXiv},
       eprint = {2107.05640},
 primaryClass = {astro-ph.GA},
       adsurl = {https://ui.adsabs.harvard.edu/abs/2022ApJ...925...74J}
}

@ARTICLE{2010AJ....140..533G,
       author = {{Griffith}, Roger L. and {Stern}, Daniel},
        title = "{Morphologies of Radio-, X-ray-, and Mid-infrared-selected Active Galactic Nuclei}",
      journal = {\aj},
         year = 2010,
        month = aug,
       volume = {140},
       number = {2},
        pages = {533-545},
          doi = {10.1088/0004-6256/140/2/533},
archivePrefix = {arXiv},
       eprint = {1006.1016},
 primaryClass = {astro-ph.CO},
       adsurl = {https://ui.adsabs.harvard.edu/abs/2010AJ....140..533G}
}

@ARTICLE{2009ApJ...696..891H,
       author = {{Hickox}, Ryan C. and {Jones}, Christine and {Forman}, William R. and {Murray}, Stephen S. and {Kochanek}, Christopher S. and {Eisenstein}, Daniel and {Jannuzi}, Buell T. and {Dey}, Arjun and {Brown}, Michael J.~I. and {Stern}, Daniel and {Eisenhardt}, Peter R. and {Gorjian}, Varoujan and {Brodwin}, Mark and {Narayan}, Ramesh and {Cool}, Richard J. and {Kenter}, Almus and {Caldwell}, Nelson and {Anderson}, Michael E.},
        title = "{Host Galaxies, Clustering, Eddington Ratios, and Evolution of Radio, X-Ray, and Infrared-Selected AGNs}",
      journal = {\apj},
         year = 2009,
        month = may,
       volume = {696},
       number = {1},
        pages = {891-919},
          doi = {10.1088/0004-637X/696/1/891},
archivePrefix = {arXiv},
       eprint = {0901.4121},
 primaryClass = {astro-ph.GA},
       adsurl = {https://ui.adsabs.harvard.edu/abs/2009ApJ...696..891H}
}

@ARTICLE{2020PASP..132e4101G,
       author = {{Gu}, Yizhou and {Fang}, Guanwen and {Yuan}, Qirong and {Lu}, Shiying},
        title = "{Structures, Stellar Population Properties, AGN Fractions, and Environments of Massive Compact Galaxies at 1 < z < 2 in 3D-HST/CANDELS}",
      journal = {\pasp},
         year = 2020,
        month = may,
       volume = {132},
       number = {1011},
          eid = {054101},
        pages = {054101},
          doi = {10.1088/1538-3873/ab797d},
archivePrefix = {arXiv},
       eprint = {2002.11274},
 primaryClass = {astro-ph.GA},
       adsurl = {https://ui.adsabs.harvard.edu/abs/2020PASP..132e4101G}
}

@ARTICLE{2014ApJ...791...52B,
       author = {{Barro}, G. and {Faber}, S.~M. and {P{\'e}rez-Gonz{\'a}lez}, P.~G. and {Pacifici}, C. and {Trump}, J.~R. and {Koo}, D.~C. and {Wuyts}, S. and {Guo}, Y. and {Bell}, E. and {Dekel}, A. and {Porter}, L. and {Primack}, J. and {Ferguson}, H. and {Ashby}, M.~L.~N. and {Caputi}, K. and {Ceverino}, D. and {Croton}, D. and {Fazio}, G.~G. and {Giavalisco}, M. and {Hsu}, L. and {Kocevski}, D. and {Koekemoer}, A. and {Kurczynski}, P. and {Kollipara}, P. and {Lee}, J. and {McIntosh}, D.~H. and {McGrath}, E. and {Moody}, C. and {Somerville}, R. and {Papovich}, C. and {Salvato}, M. and {Santini}, P. and {Tal}, T. and {van der Wel}, A. and {Williams}, C.~C. and {Willner}, S.~P. and {Zolotov}, A.},
        title = "{CANDELS+3D-HST: Compact SFGs at z \raisebox{-0.5ex}\textasciitilde 2-3, the Progenitors of the First Quiescent Galaxies}",
      journal = {\apj},
         year = 2014,
        month = aug,
       volume = {791},
       number = {1},
          eid = {52},
        pages = {52},
          doi = {10.1088/0004-637X/791/1/52},
archivePrefix = {arXiv},
       eprint = {1311.5559},
 primaryClass = {astro-ph.CO},
       adsurl = {https://ui.adsabs.harvard.edu/abs/2014ApJ...791...52B}
}

@ARTICLE{2013ApJ...765..104B,
       author = {{Barro}, Guillermo and {Faber}, S.~M. and {P{\'e}rez-Gonz{\'a}lez}, Pablo G. and {Koo}, David C. and {Williams}, Christina C. and {Kocevski}, Dale D. and {Trump}, Jonathan R. and {Mozena}, Mark and {McGrath}, Elizabeth and {van der Wel}, Arjen and {Wuyts}, Stijn and {Bell}, Eric F. and {Croton}, Darren J. and {Ceverino}, Daniel and {Dekel}, Avishai and {Ashby}, M.~L.~N. and {Cheung}, Edmond and {Ferguson}, Henry C. and {Fontana}, Adriano and {Fang}, Jerome and {Giavalisco}, Mauro and {Grogin}, Norman A. and {Guo}, Yicheng and {Hathi}, Nimish P. and {Hopkins}, Philip F. and {Huang}, Kuang-Han and {Koekemoer}, Anton M. and {Kartaltepe}, Jeyhan S. and {Lee}, Kyoung-Soo and {Newman}, Jeffrey A. and {Porter}, Lauren A. and {Primack}, Joel R. and {Ryan}, Russell E. and {Rosario}, David and {Somerville}, Rachel S. and {Salvato}, Mara and {Hsu}, Li-Ting},
        title = "{CANDELS: The Progenitors of Compact Quiescent Galaxies at z \raisebox{-0.5ex}\textasciitilde 2}",
      journal = {\apj},
         year = 2013,
        month = mar,
       volume = {765},
       number = {2},
          eid = {104},
        pages = {104},
          doi = {10.1088/0004-637X/765/2/104},
archivePrefix = {arXiv},
       eprint = {1206.5000},
 primaryClass = {astro-ph.CO},
       adsurl = {https://ui.adsabs.harvard.edu/abs/2013ApJ...765..104B}
}

@ARTICLE{2025ApJ...994..265V,
       author = {{Vijarnwannaluk}, Bovornpratch and {Gao}, Zhen-Kai and {Wang}, Wei-Hao and {Chen}, Chian-Chou and {Naufal}, Abdurrahman and {Ranjan}, Adarsh and {Hsieh}, Bau-Ching and {Mondal}, Chayan and {Chang}, Chih-Yuan and {Algera}, Hiddo S.~B. and {Liao}, Li-Wen and {Akiyama}, Masayuki and {Kim}, Seong Jin and {Mizukoshi}, Shoichiro and {Goto}, Tomotsugo and {Chang}, Yu-Yen and {Casey}, Caitlin and {Kartaltepe}, Jeyhan S. and {Akins}, Hollis B. and {Shuntov}, Marko and {Franco}, Maximilien and {Harish}, Santosh},
        title = "{The Stellar Morphology and Size of X-Ray-selected Active Galactic Nucleus Host Galaxies Revealed by JWST}",
      journal = {\apj},
         year = 2025,
        month = dec,
       volume = {994},
       number = {2},
          eid = {265},
        pages = {265},
          doi = {10.3847/1538-4357/ae102a},
archivePrefix = {arXiv},
       eprint = {2510.13719},
 primaryClass = {astro-ph.GA},
       adsurl = {https://ui.adsabs.harvard.edu/abs/2025ApJ...994..265V}
}

@ARTICLE{2021ApJ...918...22L,
       author = {{Li}, Junyao and {Silverman}, John D. and {Ding}, Xuheng and {Strauss}, Michael A. and {Goulding}, Andy and {Birrer}, Simon and {Yesuf}, Hassen M. and {Xue}, Yongquan and {Kawinwanichakij}, Lalitwadee and {Matsuoka}, Yoshiki and {Toba}, Yoshiki and {Nagao}, Tohru and {Schramm}, Malte and {Inayoshi}, Kohei},
        title = "{The Sizes of Quasar Host Galaxies in the Hyper Suprime-Cam Subaru Strategic Program}",
      journal = {\apj},
         year = 2021,
        month = sep,
       volume = {918},
       number = {1},
          eid = {22},
        pages = {22},
          doi = {10.3847/1538-4357/ac06a8},
archivePrefix = {arXiv},
       eprint = {2105.06568},
 primaryClass = {astro-ph.GA},
       adsurl = {https://ui.adsabs.harvard.edu/abs/2021ApJ...918...22L}
}

@ARTICLE{2024MNRAS.527.4690L,
       author = {{Li}, Junyao and {Silverman}, John D. and {Merloni}, Andrea and {Salvato}, Mara and {Buchner}, Johannes and {Goulding}, Andy and {Liu}, Teng and {Arcodia}, Riccardo and {Comparat}, Johan and {Ding}, Xuheng and {Ichikawa}, Kohei and {Imanishi}, Masatoshi and {Kawaguchi}, Toshihiro and {Kawinwanichakij}, Lalitwadee and {Toba}, Yoshiki},
        title = "{The eROSITA final equatorial-depth survey (eFEDS): host-galaxy demographics of X-ray AGNs with Subaru Hyper Suprime-Cam}",
      journal = {\mnras},
         year = 2024,
        month = jan,
       volume = {527},
       number = {3},
        pages = {4690-4704},
          doi = {10.1093/mnras/stad3438},
archivePrefix = {arXiv},
       eprint = {2302.12438},
 primaryClass = {astro-ph.GA},
       adsurl = {https://ui.adsabs.harvard.edu/abs/2024MNRAS.527.4690L}
}

@ARTICLE{2019ApJ...887L...5S,
       author = {{Silverman}, John D. and {Treu}, Tommaso and {Ding}, Xuheng and {Jahnke}, Knud and {Bennert}, Vardha N. and {Birrer}, Simon and {Schramm}, Malte and {Schulze}, Andreas and {Kartaltepe}, Jeyhan S. and {Sanders}, David B. and {Cen}, Renyue},
        title = "{Where Do Quasar Hosts Lie with Respect to the Size-Mass Relation of Galaxies?}",
      journal = {\apjl},
         year = 2019,
        month = dec,
       volume = {887},
       number = {1},
          eid = {L5},
        pages = {L5},
          doi = {10.3847/2041-8213/ab5851},
archivePrefix = {arXiv},
       eprint = {1910.14242},
 primaryClass = {astro-ph.GA},
       adsurl = {https://ui.adsabs.harvard.edu/abs/2019ApJ...887L...5S}
}

@ARTICLE{2026arXiv260614869Y,
       author = {{Yang}, Lilan and {Kartaltepe}, Jeyhan S. and {Drakos}, Nicole E. and {Faisst}, Andreas L. and {Flayhart}, Carter and {Franco}, Maximilien and {Koekemoer}, Anton M. and {Ilbert}, Olivier and {Akins}, Hollis B. and {Casey}, Caitlin M. and {Ding}, Xuheng and {Hadi}, Ali and {Harish}, Santosh and {Laishram}, Ronaldo and {Liu}, Daizhong and {Magdis}, Georgios E. and {Martinez}, III, Felix and {McCracken}, Henry Joy and {Paquereau}, Louise and {Rhodes}, Jason and {Robertson}, Brant E. and {Shuntov}, Marko and {Toni}, Greta},
        title = "{COSMOS-Web: A Multi-wavelength Morphological Catalog of \raisebox{-0.5ex}\textasciitilde780,000 Galaxies}",
      journal = {arXiv e-prints},
         year = 2026,
        month = jun,
          eid = {arXiv:2606.14869},
        pages = {arXiv:2606.14869},
          doi = {10.48550/arXiv.2606.14869},
archivePrefix = {arXiv},
       eprint = {2606.14869},
 primaryClass = {astro-ph.GA},
       adsurl = {https://ui.adsabs.harvard.edu/abs/2026arXiv260614869Y}
}

@ARTICLE{2026ApJS..282....6K,
       author = {{Khostovan}, Ali Ahmad and {Kartaltepe}, Jeyhan S. and {Salvato}, Mara and {Ilbert}, Olivier and {Casey}, Caitlin M. and {Algera}, Hiddo and {Antwi-Danso}, Jacqueline and {Battisti}, Andrew and {Brinch}, Malte and {Brusa}, Marcella and {Calabr{\`o}}, Antonello and {Capak}, Peter L. and {Chartab}, Nima and {Cooper}, Olivia R. and {Cox}, Isa G. and {Darvish}, Behnam and {Drakos}, Nicole E. and {Faisst}, Andreas L. and {George}, Matthew R. and {Gozaliasl}, Ghassem and {Harish}, Santosh and {Hasinger}, G{\"u}nther and {Hatamnia}, Hossein and {Iovino}, Angela and {Jin}, Shuowen and {Kashino}, Daichi and {Koekemoer}, Anton M. and {Laishram}, Ronaldo and {Lee}, Khee-Gan and {Lertprasertpong}, Jitrapon and {Lilly}, Simon J. and {Liu}, Daizhong and {Masters}, Daniel C. and {Mobasher}, Bahram and {Nagao}, Tohru and {Onodera}, Masato and {Peng}, Yingjie and {Sanders}, David B. and {Sanders}, Ryan L. and {Sattari}, Zahra and {Scoville}, Nick and {Shah}, Ekta A. and {Silverman}, John D. and {Suzuki}, Nao and {Taamoli}, Sina and {Tanaka}, Masayuki and {Tasca}, Lidia A.~M. and {Toft}, Sune and {Toni}, Greta and {Trakhtenbrot}, Benny and {Trump}, Jonathan R. and {Vaccari}, Mattia and {Valentino}, Francesco and {Vanderhoof}, Brittany N. and {Weaver}, John R. and {Yun}, Min S. and {Zavala}, Jorge A.},
        title = "{COSMOS Spectroscopic Redshift Compilation (First Data Release): 488,000 Redshifts Encompassing Two Decades of Spectroscopy}",
      journal = {\apjs},
         year = 2026,
        month = jan,
       volume = {282},
       number = {1},
          eid = {6},
        pages = {6},
          doi = {10.3847/1538-4365/ae1cb9},
archivePrefix = {arXiv},
       eprint = {2503.00120},
 primaryClass = {astro-ph.GA},
       adsurl = {https://ui.adsabs.harvard.edu/abs/2026ApJS..282....6K}
}

@ARTICLE{2017A&A...602A...3D,
       author = {{Delvecchio}, I. and {Smol{\v{c}}i{\'c}}, V. and {Zamorani}, G. and {Lagos}, C. Del P. and {Berta}, S. and {Delhaize}, J. and {Baran}, N. and {Alexander}, D.~M. and {Rosario}, D.~J. and {Gonzalez-Perez}, V. and {Ilbert}, O. and {Lacey}, C.~G. and {Le F{\`e}vre}, O. and {Miettinen}, O. and {Aravena}, M. and {Bondi}, M. and {Carilli}, C. and {Ciliegi}, P. and {Mooley}, K. and {Novak}, M. and {Schinnerer}, E. and {Capak}, P. and {Civano}, F. and {Fanidakis}, N. and {Herrera Ruiz}, N. and {Karim}, A. and {Laigle}, C. and {Marchesi}, S. and {McCracken}, H.~J. and {Middleberg}, E. and {Salvato}, M. and {Tasca}, L.},
        title = "{The VLA-COSMOS 3 GHz Large Project: AGN and host-galaxy properties out to z {\ensuremath{\lesssim}} 6}",
      journal = {\aap},
         year = 2017,
        month = jun,
       volume = {602},
          eid = {A3},
        pages = {A3},
          doi = {10.1051/0004-6361/201629367},
archivePrefix = {arXiv},
       eprint = {1703.09720},
 primaryClass = {astro-ph.GA},
       adsurl = {https://ui.adsabs.harvard.edu/abs/2017A&A...602A...3D}
}

@ARTICLE{2026AJ....171..285D,
       author = {{DESI Collaboration} and {Abdul Karim}, M. and {Adame}, A.~G. and {Aguado}, D. and {Aguilar}, J. and {Ahlen}, S. and {Alam}, S. and {Aldering}, G. and {Alexander}, D.~M. and {Alfarsy}, R. and {Allen}, L. and {Allende Prieto}, C. and {Alves}, O. and {Anand}, A. and {Andrade}, U. and {Armengaud}, E. and {Avila}, S. and {Aviles}, A. and {Awan}, H. and {Bailey}, S. and {Baleato Lizancos}, A. and {Ballester}, O. and {Bault}, A. and {Bautista}, J. and {Bean}, R. and {Behera}, J. and {BenZvi}, S. and {Beraldo e Silva}, L. and {Bermejo-Climent}, J.~R. and {Beutler}, F. and {Bianchi}, D. and {Blake}, C. and {Blum}, R. and {Bolton}, A.~S. and {Bonici}, M. and {Brieden}, S. and {Brodzeller}, A. and {Brooks}, D. and {Buckley-Geer}, E. and {Burtin}, E. and {Bystr{\"o}m}, A. and {Canning}, R. and {Carnero Rosell}, A. and {Carr}, A. and {Carrilho}, P. and {Casas}, L. and {Castander}, F.~J. and {Cereskaite}, R. and {Cervantes-Cota}, J.~L. and {Chaussidon}, E. and {Chaves-Montero}, J. and {Chen}, S. and {Chen}, X. and {Circosta}, C. and {Claybaugh}, T. and {Cole}, S. and {Cooper}, A.~P. and {Cousinou}, M.-C. and {Cuceu}, A. and {Davis}, T.~M. and {Dawson}, K.~S. and {de Belsunce}, R. and {de la Cruz}, R. and {de la Macorra}, A. and {de Mattia}, A. and {Deiosso}, N. and {Della Costa}, J. and {Demina}, R. and {Demirbozan}, U. and {DeRose}, J. and {Dey}, A. and {Dey}, B. and {Ding}, J. and {Ding}, Z. and {Doel}, P. and {Douglass}, K. and {Dowicz}, M. and {Ebina}, H. and {Edelstein}, J. and {Eisenstein}, D.~J. and {Elbers}, W. and {Emas}, N. and {Escoffier}, S. and {Fagrelius}, P. and {Fan}, X. and {Fanning}, K. and {Favole}, G. and {Fawcett}, V.~A. and {Fern{\'a}ndez-Garc{\'\i}a}, E. and {Ferraro}, S. and {Findlay}, N. and {Font-Ribera}, A. and {Forero-Romero}, J.~E. and {Forero-S{\'a}nchez}, D. and {Frenk}, C.~S. and {G{\"a}nsicke}, B.~T. and {Galbany}, L. and {Garc{\'\i}a-Bellido}, J. and {Garcia-Quintero}, C. and {Garrison}, L.~H. and {Gazta{\~n}aga}, E. and {Gil-Mar{\'\i}n}, H. and {Gloudemans}, A. and {Gnedin}, O.~Y. and {Gontcho A Gontcho}, S. and {Gonzalez}, D. and {Gonzalez-Morales}, A.~X. and {Gonzalez-Perez}, V. and {Gordon}, C. and {Graur}, O. and {Green}, D. and {Gruen}, D. and {Gsponer}, R. and {Guandalin}, C. and {Gutierrez}, G. and {Guy}, J. and {Hahn}, C. and {Han}, J.~J. and {Han}, J. and {He}, S. and {Herrera-Alcantar}, H.~K. and {Heydenreich}, S. and {Honscheid}, K. and {Hou}, J. and {Howlett}, C. and {Huterer}, D. and {Ir{\v{s}}i{\v{c}}}, V. and {Ishak}, M. and {Jacques}, A. and {Jiang}, L. and {Jimenez}, J. and {Jing}, Y.~P. and {Joachimi}, B. and {Joudaki}, S. and {Joyce}, R. and {Jullo}, E. and {Juneau}, S. and {Kara{\c{c}}ayl{\i}}, N.~G. and {Karim}, T. and {Kehoe}, R. and {Kent}, S. and {Khederlarian}, A. and {Kirkby}, D. and {Kisner}, T. and {Kitaura}, F.-S. and {Kizhuprakkat}, N. and {Kong}, H. and {Koposov}, S.~E. and {Kremin}, A. and {Krolewski}, A. and {Lahav}, O. and {Lai}, Y. and {Lamman}, C. and {Lan}, T.-W. and {Landriau}, M. and {Lang}, D. and {Lange}, J.~U. and {Lasker}, J. and {Le Goff}, J.~M. and {Le Guillou}, L. and {Leauthaud}, A. and {Levi}, M.~E. and {Li}, S. and {Li}, T.~S. and {Liu}, W. and {Lodha}, K. and {Lokken}, M. and {Luo}, Y. and {Magneville}, C. and {Manera}, M. and {Manser}, C.~J. and {Margala}, D. and {Martini}, P. and {Maus}, M. and {McCullough}, J. and {McDonald}, P. and {Medina}, G.~E. and {Medina-Varela}, L. and {Meisner}, A. and {Mena-Fern{\'a}ndez}, J. and {Menegas}, A. and {Meneses-Rizo}, J. and {Mezcua}, M. and {Miquel}, R. and {Montero-Camacho}, P. and {Moon}, J. and {Moustakas}, J. and {Mu{\~n}oz-Guti{\'e}rrez}, A. and {Mu noz-Santos}, D. and {Myers}, A.~D. and {Myles}, J. and {Nadathur}, S. and {Najita}, J. and {Napolitano}, L. and {Newman}, J.~A. and {Nikakhtar}, F. and {Nikutta}, R. and {Niz}, G. and {Noriega}, H.~E. and {Nugent}, P.},
        title = "{Data Release 1 of the Dark Energy Spectroscopic Instrument}",
      journal = {\aj},
         year = 2026,
        month = may,
       volume = {171},
       number = {5},
          eid = {285},
        pages = {285},
          doi = {10.3847/1538-3881/ae4c43},
archivePrefix = {arXiv},
       eprint = {2503.14745},
 primaryClass = {astro-ph.CO},
       adsurl = {https://ui.adsabs.harvard.edu/abs/2026AJ....171..285D}
}

@ARTICLE{2022arXiv221202428K,
       author = {{K{\"u}mmel}, M. and {{\'A}lvarez-Ayll{\'o}n}, A. and {Bertin}, E. and {Dubath}, P. and {Gavazzi}, R. and {Hartley}, W. and {Schefer}, M.},
        title = "{Using the SourceXtractor++ package for data reduction}",
      journal = {arXiv e-prints},
         year = 2022,
        month = dec,
          eid = {arXiv:2212.02428},
        pages = {arXiv:2212.02428},
          doi = {10.48550/arXiv.2212.02428},
archivePrefix = {arXiv},
       eprint = {2212.02428},
 primaryClass = {astro-ph.IM},
       adsurl = {https://ui.adsabs.harvard.edu/abs/2022arXiv221202428K}
}

@INPROCEEDINGS{2020ASPC..527...29K,
       author = {{K{\"u}mmel}, M. and {Bertin}, E. and {Schefer}, M. and {Apostolakos}, N. and {{\'A}lvarez-Ayll{\'o}n}, A. and {Dubath}, P.},
        title = "{Working With the SourceXtractor++ Software}",
    booktitle = {Astronomical Data Analysis Software and Systems XXIX},
         year = 2020,
       editor = {{Pizzo}, R. and {Deul}, E.~R. and {Mol}, J.~D. and {de Plaa}, J. and {Verkouter}, H.},
       series = {Astronomical Society of the Pacific Conference Series},
       volume = {527},
        month = jan,
        pages = {29},
       adsurl = {https://ui.adsabs.harvard.edu/abs/2020ASPC..527...29K}
}

@INPROCEEDINGS{2020ASPC..527..461B,
       author = {{Bertin}, E. and {Schefer}, M. and {Apostolakos}, N. and {{\'A}lvarez-Ayll{\'o}n}, A. and {Dubath}, P. and {K{\"u}mmel}, M.},
        title = "{The SourceXtractor++ Software}",
    booktitle = {Astronomical Data Analysis Software and Systems XXIX},
         year = 2020,
       editor = {{Pizzo}, R. and {Deul}, E.~R. and {Mol}, J.~D. and {de Plaa}, J. and {Verkouter}, H.},
       series = {Astronomical Society of the Pacific Conference Series},
       volume = {527},
        month = jan,
        pages = {461},
       adsurl = {https://ui.adsabs.harvard.edu/abs/2020ASPC..527..461B}
}

@ARTICLE{2019A&A...622A.103B,
       author = {{Boquien}, M. and {Burgarella}, D. and {Roehlly}, Y. and {Buat}, V. and {Ciesla}, L. and {Corre}, D. and {Inoue}, A.~K. and {Salas}, H.},
        title = "{CIGALE: a python Code Investigating GALaxy Emission}",
      journal = {\aap},
         year = 2019,
        month = feb,
       volume = {622},
          eid = {A103},
        pages = {A103},
          doi = {10.1051/0004-6361/201834156},
archivePrefix = {arXiv},
       eprint = {1811.03094},
 primaryClass = {astro-ph.GA},
       adsurl = {https://ui.adsabs.harvard.edu/abs/2019A&A...622A.103B}
}

@ARTICLE{2006A&A...457..841I,
       author = {{Ilbert}, O. and {Arnouts}, S. and {McCracken}, H.~J. and {Bolzonella}, M. and {Bertin}, E. and {Le F{\`e}vre}, O. and {Mellier}, Y. and {Zamorani}, G. and {Pell{\`o}}, R. and {Iovino}, A. and {Tresse}, L. and {Le Brun}, V. and {Bottini}, D. and {Garilli}, B. and {Maccagni}, D. and {Picat}, J.~P. and {Scaramella}, R. and {Scodeggio}, M. and {Vettolani}, G. and {Zanichelli}, A. and {Adami}, C. and {Bardelli}, S. and {Cappi}, A. and {Charlot}, S. and {Ciliegi}, P. and {Contini}, T. and {Cucciati}, O. and {Foucaud}, S. and {Franzetti}, P. and {Gavignaud}, I. and {Guzzo}, L. and {Marano}, B. and {Marinoni}, C. and {Mazure}, A. and {Meneux}, B. and {Merighi}, R. and {Paltani}, S. and {Pollo}, A. and {Pozzetti}, L. and {Radovich}, M. and {Zucca}, E. and {Bondi}, M. and {Bongiorno}, A. and {Busarello}, G. and {de La Torre}, S. and {Gregorini}, L. and {Lamareille}, F. and {Mathez}, G. and {Merluzzi}, P. and {Ripepi}, V. and {Rizzo}, D. and {Vergani}, D.},
        title = "{Accurate photometric redshifts for the CFHT legacy survey calibrated using the VIMOS VLT deep survey}",
      journal = {\aap},
         year = 2006,
        month = oct,
       volume = {457},
       number = {3},
        pages = {841-856},
          doi = {10.1051/0004-6361:20065138},
archivePrefix = {arXiv},
       eprint = {astro-ph/0603217},
 primaryClass = {astro-ph},
       adsurl = {https://ui.adsabs.harvard.edu/abs/2006A&A...457..841I}
}

@ARTICLE{2002MNRAS.329..355A,
       author = {{Arnouts}, S. and {Moscardini}, L. and {Vanzella}, E. and {Colombi}, S. and {Cristiani}, S. and {Fontana}, A. and {Giallongo}, E. and {Matarrese}, S. and {Saracco}, P.},
        title = "{Measuring the redshift evolution of clustering: the Hubble Deep Field South}",
      journal = {\mnras},
         year = 2002,
        month = jan,
       volume = {329},
       number = {2},
        pages = {355-366},
          doi = {10.1046/j.1365-8711.2002.04988.x},
archivePrefix = {arXiv},
       eprint = {astro-ph/0109453},
 primaryClass = {astro-ph},
       adsurl = {https://ui.adsabs.harvard.edu/abs/2002MNRAS.329..355A}
}

@ARTICLE{2003PASP..115..763C,
       author = {{Chabrier}, Gilles},
        title = "{Galactic Stellar and Substellar Initial Mass Function}",
      journal = {\pasp},
         year = 2003,
        month = jul,
       volume = {115},
       number = {809},
        pages = {763-795},
          doi = {10.1086/376392},
archivePrefix = {arXiv},
       eprint = {astro-ph/0304382},
 primaryClass = {astro-ph},
       adsurl = {https://ui.adsabs.harvard.edu/abs/2003PASP..115..763C}
}

@ARTICLE{2011MNRAS.415.2920I,
       author = {{Inoue}, Akio K.},
        title = "{Rest-frame ultraviolet-to-optical spectral characteristics of extremely metal-poor and metal-free galaxies}",
      journal = {\mnras},
         year = 2011,
        month = aug,
       volume = {415},
       number = {3},
        pages = {2920-2931},
          doi = {10.1111/j.1365-2966.2011.18906.x},
archivePrefix = {arXiv},
       eprint = {1102.5150},
 primaryClass = {astro-ph.CO},
       adsurl = {https://ui.adsabs.harvard.edu/abs/2011MNRAS.415.2920I}
}

@ARTICLE{2012MNRAS.420.2756S,
       author = {{Stalevski}, Marko and {Fritz}, Jacopo and {Baes}, Maarten and {Nakos}, Theodoros and {Popovi{\'c}}, Luka {\v{C}}.},
        title = "{3D radiative transfer modelling of the dusty tori around active galactic nuclei as a clumpy two-phase medium}",
      journal = {\mnras},
         year = 2012,
        month = mar,
       volume = {420},
       number = {4},
        pages = {2756-2772},
          doi = {10.1111/j.1365-2966.2011.19775.x},
archivePrefix = {arXiv},
       eprint = {1109.1286},
 primaryClass = {astro-ph.CO},
       adsurl = {https://ui.adsabs.harvard.edu/abs/2012MNRAS.420.2756S}
}

@ARTICLE{2003MNRAS.344.1000B,
       author = {{Bruzual}, G. and {Charlot}, S.},
        title = "{Stellar population synthesis at the resolution of 2003}",
      journal = {\mnras},
         year = 2003,
        month = oct,
       volume = {344},
       number = {4},
        pages = {1000-1028},
          doi = {10.1046/j.1365-8711.2003.06897.x},
archivePrefix = {arXiv},
       eprint = {astro-ph/0309134},
 primaryClass = {astro-ph},
       adsurl = {https://ui.adsabs.harvard.edu/abs/2003MNRAS.344.1000B}
}

@ARTICLE{2006MNRAS.365..807M,
       author = {{Meiksin}, Avery},
        title = "{Colour corrections for high-redshift objects due to intergalactic attenuation}",
      journal = {\mnras},
         year = 2006,
        month = jan,
       volume = {365},
       number = {3},
        pages = {807-812},
          doi = {10.1111/j.1365-2966.2005.09756.x},
archivePrefix = {arXiv},
       eprint = {astro-ph/0512435},
 primaryClass = {astro-ph},
       adsurl = {https://ui.adsabs.harvard.edu/abs/2006MNRAS.365..807M}
}

@ARTICLE{2016MNRAS.458.2288S,
       author = {{Stalevski}, Marko and {Ricci}, Claudio and {Ueda}, Yoshihiro and {Lira}, Paulina and {Fritz}, Jacopo and {Baes}, Maarten},
        title = "{The dust covering factor in active galactic nuclei}",
      journal = {\mnras},
         year = 2016,
        month = may,
       volume = {458},
       number = {3},
        pages = {2288-2302},
          doi = {10.1093/mnras/stw444},
archivePrefix = {arXiv},
       eprint = {1602.06954},
 primaryClass = {astro-ph.GA},
       adsurl = {https://ui.adsabs.harvard.edu/abs/2016MNRAS.458.2288S}
}

@ARTICLE{2014ApJ...784...83D,
       author = {{Dale}, Daniel A. and {Helou}, George and {Magdis}, Georgios E. and {Armus}, Lee and {D{\'\i}az-Santos}, Tanio and {Shi}, Yong},
        title = "{A Two-parameter Model for the Infrared/Submillimeter/Radio Spectral Energy Distributions of Galaxies and Active Galactic Nuclei}",
      journal = {\apj},
         year = 2014,
        month = mar,
       volume = {784},
       number = {1},
          eid = {83},
        pages = {83},
          doi = {10.1088/0004-637X/784/1/83},
archivePrefix = {arXiv},
       eprint = {1402.1495},
 primaryClass = {astro-ph.GA},
       adsurl = {https://ui.adsabs.harvard.edu/abs/2014ApJ...784...83D}
}

@ARTICLE{2002ApJS..140..303L,
       author = {{Leitherer}, Claus and {Li}, I.-Hui and {Calzetti}, Daniela and {Heckman}, Timothy M.},
        title = "{Global Far-Ultraviolet (912-1800 {\r{A}}) Properties of Star-forming Galaxies}",
      journal = {\apjs},
         year = 2002,
        month = jun,
       volume = {140},
       number = {2},
        pages = {303-329},
          doi = {10.1086/342486},
       adsurl = {https://ui.adsabs.harvard.edu/abs/2002ApJS..140..303L}
}

@ARTICLE{2000ApJ...533..682C,
       author = {{Calzetti}, Daniela and {Armus}, Lee and {Bohlin}, Ralph C. and {Kinney}, Anne L. and {Koornneef}, Jan and {Storchi-Bergmann}, Thaisa},
        title = "{The Dust Content and Opacity of Actively Star-forming Galaxies}",
      journal = {\apj},
         year = 2000,
        month = apr,
       volume = {533},
       number = {2},
        pages = {682-695},
          doi = {10.1086/308692},
archivePrefix = {arXiv},
       eprint = {astro-ph/9911459},
 primaryClass = {astro-ph},
       adsurl = {https://ui.adsabs.harvard.edu/abs/2000ApJ...533..682C}
}

@ARTICLE{2020NaMet..17..261V,
       author = {{Virtanen}, Pauli and {Gommers}, Ralf and {Oliphant}, Travis E. and {Haberland}, Matt and {Reddy}, Tyler and {Cournapeau}, David and {Burovski}, Evgeni and {Peterson}, Pearu and {Weckesser}, Warren and {Bright}, Jonathan and {van der Walt}, St{\'e}fan J. and {Brett}, Matthew and {Wilson}, Joshua and {Millman}, K. Jarrod and {Mayorov}, Nikolay and {Nelson}, Andrew R.~J. and {Jones}, Eric and {Kern}, Robert and {Larson}, Eric and {Carey}, C.~J. and {Polat}, {\.I}lhan and {Feng}, Yu and {Moore}, Eric W. and {VanderPlas}, Jake and {Laxalde}, Denis and {Perktold}, Josef and {Cimrman}, Robert and {Henriksen}, Ian and {Quintero}, E.~A. and {Harris}, Charles R. and {Archibald}, Anne M. and {Ribeiro}, Ant{\^o}nio H. and {Pedregosa}, Fabian and {van Mulbregt}, Paul and {SciPy 1.  0 Contributors}},
        title = "{SciPy 1.0: fundamental algorithms for scientific computing in Python}",
      journal = {Nature Methods},
         year = 2020,
        month = feb,
       volume = {17},
        pages = {261-272},
          doi = {10.1038/s41592-019-0686-2},
archivePrefix = {arXiv},
       eprint = {1907.10121},
 primaryClass = {cs.MS},
       adsurl = {https://ui.adsabs.harvard.edu/abs/2020NaMet..17..261V}
}

@INPROCEEDINGS{2005ASPC..347...29T,
       author = {{Taylor}, M.~B.},
        title = "{TOPCAT \& STIL: Starlink Table/VOTable Processing Software}",
    booktitle = {Astronomical Data Analysis Software and Systems XIV},
         year = 2005,
       editor = {{Shopbell}, P. and {Britton}, M. and {Ebert}, R.},
       series = {Astronomical Society of the Pacific Conference Series},
       volume = {347},
        month = dec,
        pages = {29},
       adsurl = {https://ui.adsabs.harvard.edu/abs/2005ASPC..347...29T}
}

@ARTICLE{2020ApJ...888...37D,
       author = {{Ding}, Xuheng and {Silverman}, John and {Treu}, Tommaso and {Schulze}, Andreas and {Schramm}, Malte and {Birrer}, Simon and {Park}, Daeseong and {Jahnke}, Knud and {Bennert}, Vardha N. and {Kartaltepe}, Jeyhan S. and {Koekemoer}, Anton M. and {Malkan}, Matthew A. and {Sanders}, David},
        title = "{The Mass Relations between Supermassive Black Holes and Their Host Galaxies at 1 < z < 2 HST-WFC3}",
      journal = {\apj},
         year = 2020,
        month = jan,
       volume = {888},
       number = {1},
          eid = {37},
        pages = {37},
          doi = {10.3847/1538-4357/ab5b90},
archivePrefix = {arXiv},
       eprint = {1910.11875},
 primaryClass = {astro-ph.GA},
       adsurl = {https://ui.adsabs.harvard.edu/abs/2020ApJ...888...37D}
}

@ARTICLE{2025A&A...704A.339S,
       author = {{Shuntov}, Marko and {Akins}, Hollis B. and {Paquereau}, Louise and {Casey}, Caitlin M. and {Ilbert}, Olivier and {Arango-Toro}, Rafael C. and {McCracken}, Henry Joy and {Franco}, Maximilien and {Harish}, Santosh and {Kartaltepe}, Jeyhan S. and {Koekemoer}, Anton M. and {Yang}, Lilan and {Huertas-Company}, Marc and {Berman}, Edward M. and {McCleary}, Jacqueline E. and {Toft}, Sune and {Gavazzi}, Rapha{\"e}l and {Achenbach}, Mark J. and {Bertin}, Emmanuel and {Brinch}, Malte and {Champagne}, Jackie and {Chartab}, Nima and {Drakos}, Nicole E. and {Egami}, Eiichi and {Endsley}, Ryan and {Faisst}, Andreas L. and {Fan}, Xiaohui and {Flayhart}, Carter and {Hartley}, William G. and {Hatamnia}, Hossein and {Gozaliasl}, Ghassem and {Gentile}, Fabrizio and {Jermann}, Iris and {Jin}, Shuowen and {Kakiichi}, Koki and {Khostovan}, Ali Ahmad and {K{\"u}mmel}, Martin and {Laigle}, Clotilde and {Laishram}, Ronaldo and {Lambrides}, Erini and {Liu}, Daizhong and {Lyu}, Jianwei and {Magdis}, Georgios and {Mobasher}, Bahram and {Moutard}, Thibaud and {Renzini}, Alvio and {Rich}, R. Michael and {Sanders}, David B. and {Sattari}, Zahra and {Robertson}, Brant E. and {Schefer}, Marc and {Scognamiglio}, Diana and {Scoville}, Nick and {Silverman}, John D. and {Taamoli}, Sina and {Trakhtenbrot}, Benny and {Valentino}, Francesco and {Wang}, Feige and {Weaver}, John R. and {Yang}, Jinyi},
        title = "{COSMOS2025: The COSMOS-Web galaxy catalog of photometry, morphology, redshifts, and physical parameters from JWST, HST, and ground-based imaging}",
      journal = {\aap},
         year = 2025,
        month = dec,
       volume = {704},
          eid = {A339},
        pages = {A339},
          doi = {10.1051/0004-6361/202555799},
archivePrefix = {arXiv},
       eprint = {2506.03243},
 primaryClass = {astro-ph.GA},
       adsurl = {https://ui.adsabs.harvard.edu/abs/2025A&A...704A.339S}
}

@ARTICLE{2018MNRAS.480.4379C,
       author = {{Carnall}, A.~C. and {McLure}, R.~J. and {Dunlop}, J.~S. and {Dav{\'e}}, R.},
        title = "{Inferring the star formation histories of massive quiescent galaxies with BAGPIPES: evidence for multiple quenching mechanisms}",
      journal = {\mnras},
         year = 2018,
        month = nov,
       volume = {480},
       number = {4},
        pages = {4379-4401},
          doi = {10.1093/mnras/sty2169},
archivePrefix = {arXiv},
       eprint = {1712.04452},
 primaryClass = {astro-ph.GA},
       adsurl = {https://ui.adsabs.harvard.edu/abs/2018MNRAS.480.4379C}
}

@ARTICLE{2005ApJ...621..587C,
       author = {{Colbert}, James W. and {Teplitz}, Harry I. and {Yan}, Lin and {Malkan}, Matthew A. and {McCarthy}, Patrick J.},
        title = "{Near-Infrared Properties of Faint X-Ray Sources from NICMOS Imaging in the Chandra Deep Fields}",
      journal = {\apj},
         year = 2005,
        month = mar,
       volume = {621},
       number = {2},
        pages = {587-595},
          doi = {10.1086/426960},
archivePrefix = {arXiv},
       eprint = {astro-ph/0411433},
 primaryClass = {astro-ph},
       adsurl = {https://ui.adsabs.harvard.edu/abs/2005ApJ...621..587C}
}

@ARTICLE{2026AJ....171...71R,
       author = {{Ratajczak}, J. and {Dawson}, K.~S. and {Weaverdyck}, N. and {Aguilar}, J. and {Ahlen}, S. and {Armengaud}, E. and {Bailey}, S. and {Bianchi}, D. and {Blanco}, D. and {Brodzeller}, A. and {Brooks}, D. and {Castander}, F.~J. and {Claybaugh}, T. and {Cuceu}, A. and {de la Macorra}, A. and {Dey}, Arjun and {Dey}, Biprateep and {Doel}, P. and {Font-Ribera}, A. and {Forero-Romero}, J.~E. and {Gazta{\~n}aga}, E. and {Gontcho A Gontcho}, S. and {Gutierrez}, G. and {Guy}, J. and {Hagen}, T. and {Herrera-Alcantar}, H.~K. and {Honscheid}, K. and {Huterer}, D. and {Ishak}, M. and {Jimenez}, J. and {Joyce}, R. and {Juneau}, S. and {Kehoe}, R. and {Kirkby}, D. and {Kisner}, T. and {Koposov}, S.~E. and {Kremin}, A. and {Lahav}, O. and {Lambert}, A. and {Lamman}, C. and {Landriau}, M. and {Le Guillou}, L. and {Leauthaud}, A. and {Lee}, J. and {Levi}, M.~E. and {Li}, Q. and {Longhurst}, I. and {Luo}, Y. and {Manera}, M. and {Martini}, P. and {McCullough}, J. and {Meisner}, A. and {Miquel}, R. and {Moustakas}, J. and {Nadathur}, S. and {Newman}, J.~A. and {Palanque-Delabrouille}, N. and {Percival}, W.~J. and {Poppett}, C. and {Prada}, F. and {P{\'e}rez-R{\`a}fols}, I. and {Raichoor}, A. and {Ravoux}, C. and {Rossi}, G. and {Salcedo Hernandez}, Y. and {Sanchez}, E. and {Saulder}, C. and {Schlegel}, D. and {Schubnell}, M. and {Seo}, H. and {Sprayberry}, D. and {Tarl{\'e}}, G. and {Weaver}, B.~A. and {Wechsler}, R.~H. and {White}, M. and {Zhou}, R. and {Zou}, H.},
        title = "{The Compilation and Validation of the Spectroscopic Redshift Catalogs for the DESI-COSMOS and DESI-XMM-LSS Fields}",
      journal = {\aj},
         year = 2026,
        month = feb,
       volume = {171},
       number = {2},
          eid = {71},
        pages = {71},
          doi = {10.3847/1538-3881/ae1fde},
archivePrefix = {arXiv},
       eprint = {2508.09286},
 primaryClass = {astro-ph.CO},
       adsurl = {https://ui.adsabs.harvard.edu/abs/2026AJ....171...71R}
}

@ARTICLE{1996ApJ...473..130R,
       author = {{Rush}, Brian and {Malkan}, Matthew A. and {Edelson}, Richard A.},
        title = "{The Radio Properties of Seyfert Galaxies in the 12 Micron and CfA Samples}",
      journal = {\apj},
         year = 1996,
        month = dec,
       volume = {473},
        pages = {130},
          doi = {10.1086/178132},
archivePrefix = {arXiv},
       eprint = {astro-ph/9606178},
 primaryClass = {astro-ph},
       adsurl = {https://ui.adsabs.harvard.edu/abs/1996ApJ...473..130R}
}

@ARTICLE{2023NatAs...7..463R,
       author = {{Raimundo}, Sandra I. and {Malkan}, Matthew and {Vestergaard}, Marianne},
        title = "{An increase in black hole activity in galaxies with kinematically misaligned gas}",
      journal = {Nature Astronomy},
         year = 2023,
        month = apr,
       volume = {7},
        pages = {463-472},
          doi = {10.1038/s41550-022-01880-z},
archivePrefix = {arXiv},
       eprint = {2301.07155},
 primaryClass = {astro-ph.GA},
       adsurl = {https://ui.adsabs.harvard.edu/abs/2023NatAs...7..463R}
}

@ARTICLE{2001MNRAS.322..231K,
       author = {{Kroupa}, Pavel},
        title = "{On the variation of the initial mass function}",
      journal = {\mnras},
         year = 2001,
        month = apr,
       volume = {322},
       number = {2},
        pages = {231-246},
          doi = {10.1046/j.1365-8711.2001.04022.x},
archivePrefix = {arXiv},
       eprint = {astro-ph/0009005},
 primaryClass = {astro-ph},
       adsurl = {https://ui.adsabs.harvard.edu/abs/2001MNRAS.322..231K}
}

@ARTICLE{2014ApJ...788...28V,
       author = {{van der Wel}, A. and {Franx}, M. and {van Dokkum}, P.~G. and {Skelton}, R.~E. and {Momcheva}, I.~G. and {Whitaker}, K.~E. and {Brammer}, G.~B. and {Bell}, E.~F. and {Rix}, H.-W. and {Wuyts}, S. and {Ferguson}, H.~C. and {Holden}, B.~P. and {Barro}, G. and {Koekemoer}, A.~M. and {Chang}, Yu-Yen and {McGrath}, E.~J. and {H{\"a}ussler}, B. and {Dekel}, A. and {Behroozi}, P. and {Fumagalli}, M. and {Leja}, J. and {Lundgren}, B.~F. and {Maseda}, M.~V. and {Nelson}, E.~J. and {Wake}, D.~A. and {Patel}, S.~G. and {Labb{\'e}}, I. and {Faber}, S.~M. and {Grogin}, N.~A. and {Kocevski}, D.~D.},
        title = "{3D-HST+CANDELS: The Evolution of the Galaxy Size-Mass Distribution since z = 3}",
      journal = {\apj},
         year = 2014,
        month = jun,
       volume = {788},
       number = {1},
          eid = {28},
        pages = {28},
          doi = {10.1088/0004-637X/788/1/28},
archivePrefix = {arXiv},
       eprint = {1404.2844},
 primaryClass = {astro-ph.GA},
       adsurl = {https://ui.adsabs.harvard.edu/abs/2014ApJ...788...28V}
}

@ARTICLE{2025ApJS..281...68Y,
       author = {{Yang}, Lilan and {Kartaltepe}, Jeyhan S. and {Franco}, Maximilien and {Ding}, Xuheng and {Achenbach}, Mark J. and {Arango-Toro}, Rafael C. and {Casey}, Caitlin M. and {Drakos}, Nicole E. and {Faisst}, Andreas L. and {Gillman}, Steven and {Gozaliasl}, Ghassem and {Huertas-Company}, Marc and {Jin}, Shuowen and {Liu}, Daizhong and {Magdis}, Georgios and {Massey}, Richard and {Silverman}, John D. and {Tanaka}, Takumi S. and {Yu}, Si-Yue and {Akins}, Hollis B. and {Allen}, Natalie and {Ilbert}, Olivier and {Koekemoer}, Anton M. and {McCracken}, Henry Joy and {Paquereau}, Louise and {Rhodes}, Jason and {Robertson}, Brant E. and {Shuntov}, Marko and {Toft}, Sune},
        title = "{COSMOS-Web: Unraveling the Evolution of Galaxy Size and Related Properties at 2 < z < 10}",
      journal = {\apjs},
         year = 2025,
        month = dec,
       volume = {281},
       number = {2},
          eid = {68},
        pages = {68},
          doi = {10.3847/1538-4365/ae0e1b},
archivePrefix = {arXiv},
       eprint = {2504.07185},
 primaryClass = {astro-ph.GA},
       adsurl = {https://ui.adsabs.harvard.edu/abs/2025ApJS..281...68Y}
}

@ARTICLE{2018ARA&A..56..625H,
       author = {{Hickox}, Ryan C. and {Alexander}, David M.},
        title = "{Obscured Active Galactic Nuclei}",
      journal = {\araa},
         year = 2018,
        month = sep,
       volume = {56},
        pages = {625-671},
          doi = {10.1146/annurev-astro-081817-051803},
archivePrefix = {arXiv},
       eprint = {1806.04680},
 primaryClass = {astro-ph.GA},
       adsurl = {https://ui.adsabs.harvard.edu/abs/2018ARA&A..56..625H}
}

@ARTICLE{2016MNRAS.458..242T,
       author = {{Tacchella}, Sandro and {Dekel}, Avishai and {Carollo}, C. Marcella and {Ceverino}, Daniel and {DeGraf}, Colin and {Lapiner}, Sharon and {Mandelker}, Nir and {Primack}, Joel R.},
        title = "{Evolution of density profiles in high-z galaxies: compaction and quenching inside-out}",
      journal = {\mnras},
         year = 2016,
        month = may,
       volume = {458},
       number = {1},
        pages = {242-263},
          doi = {10.1093/mnras/stw303},
archivePrefix = {arXiv},
       eprint = {1509.00017},
 primaryClass = {astro-ph.GA},
       adsurl = {https://ui.adsabs.harvard.edu/abs/2016MNRAS.458..242T}
}

@ARTICLE{2015MNRAS.450.2327Z,
       author = {{Zolotov}, Adi and {Dekel}, Avishai and {Mandelker}, Nir and {Tweed}, Dylan and {Inoue}, Shigeki and {DeGraf}, Colin and {Ceverino}, Daniel and {Primack}, Joel R. and {Barro}, Guillermo and {Faber}, Sandra M.},
        title = "{Compaction and quenching of high-z galaxies in cosmological simulations: blue and red nuggets}",
      journal = {\mnras},
         year = 2015,
        month = jul,
       volume = {450},
       number = {3},
        pages = {2327-2353},
          doi = {10.1093/mnras/stv740},
archivePrefix = {arXiv},
       eprint = {1412.4783},
 primaryClass = {astro-ph.GA},
       adsurl = {https://ui.adsabs.harvard.edu/abs/2015MNRAS.450.2327Z}
}

@ARTICLE{2012ARA&A..50..455F,
       author = {{Fabian}, A.~C.},
        title = "{Observational Evidence of Active Galactic Nuclei Feedback}",
      journal = {\araa},
         year = 2012,
        month = sep,
       volume = {50},
        pages = {455-489},
          doi = {10.1146/annurev-astro-081811-125521},
archivePrefix = {arXiv},
       eprint = {1204.4114},
 primaryClass = {astro-ph.CO},
       adsurl = {https://ui.adsabs.harvard.edu/abs/2012ARA&A..50..455F}
}

@ARTICLE{2014ARA&A..52..589H,
       author = {{Heckman}, Timothy M. and {Best}, Philip N.},
        title = "{The Coevolution of Galaxies and Supermassive Black Holes: Insights from Surveys of the Contemporary Universe}",
      journal = {\araa},
         year = 2014,
        month = aug,
       volume = {52},
        pages = {589-660},
          doi = {10.1146/annurev-astro-081913-035722},
archivePrefix = {arXiv},
       eprint = {1403.4620},
 primaryClass = {astro-ph.GA},
       adsurl = {https://ui.adsabs.harvard.edu/abs/2014ARA&A..52..589H}
}

@ARTICLE{2026ApJ..1002L..40C,
       author = {{Chen}, Longyue and {Wang}, Tao and {Sun}, Hanwen and {Xu}, Ke and {Zhou}, Luwenjia and {Yang}, Tiancheng and {Tarrasse}, Maxime and {Mo}, Houjun and {Li}, Zhaozhou and {Chen}, Yangyao and {Dekel}, Avishai and {Daddi}, Emanuele and {Ding}, Xuheng and {Giavalisco}, Mauro and {Elbaz}, David},
        title = "{A Bending in the Size─Mass Relation of Star-forming Galaxies across 0.5 < z < 6.0 at a Critical Stellar Mass of {}10$^{10}$M$_{{\ensuremath{\odot}}}$ Revealed by JWST}",
      journal = {\apjl},
         year = 2026,
        month = may,
       volume = {1002},
       number = {2},
          eid = {L40},
        pages = {L40},
          doi = {10.3847/2041-8213/ae5a93},
archivePrefix = {arXiv},
       eprint = {2603.22239},
 primaryClass = {astro-ph.GA},
       adsurl = {https://ui.adsabs.harvard.edu/abs/2026ApJ..1002L..40C}
}

@ARTICLE{2016ApJ...817...34M,
       author = {{Marchesi}, S. and {Civano}, F. and {Elvis}, M. and {Salvato}, M. and {Brusa}, M. and {Comastri}, A. and {Gilli}, R. and {Hasinger}, G. and {Lanzuisi}, G. and {Miyaji}, T. and {Treister}, E. and {Urry}, C.~M. and {Vignali}, C. and {Zamorani}, G. and {Allevato}, V. and {Cappelluti}, N. and {Cardamone}, C. and {Finoguenov}, A. and {Griffiths}, R.~E. and {Karim}, A. and {Laigle}, C. and {LaMassa}, S.~M. and {Jahnke}, K. and {Ranalli}, P. and {Schawinski}, K. and {Schinnerer}, E. and {Silverman}, J.~D. and {Smolcic}, V. and {Suh}, H. and {Trakhtenbrot}, B.},
        title = "{The Chandra COSMOS Legacy survey: optical/IR identifications}",
      journal = {\apj},
         year = 2016,
        month = jan,
       volume = {817},
       number = {1},
          eid = {34},
        pages = {34},
          doi = {10.3847/0004-637X/817/1/34},
archivePrefix = {arXiv},
       eprint = {1512.01105},
 primaryClass = {astro-ph.GA},
       adsurl = {https://ui.adsabs.harvard.edu/abs/2016ApJ...817...34M}
}

@ARTICLE{2017A&A...602A...6S,
       author = {{Smol{\v{c}}i{\'c}}, V. and {Novak}, M. and {Delvecchio}, I. and {Ceraj}, L. and {Bondi}, M. and {Delhaize}, J. and {Marchesi}, S. and {Murphy}, E. and {Schinnerer}, E. and {Vardoulaki}, E. and {Zamorani}, G.},
        title = "{The VLA-COSMOS 3 GHz Large Project: Cosmic evolution of radio AGN and implications for radio-mode feedback since z   5}",
      journal = {\aap},
         year = 2017,
        month = jun,
       volume = {602},
          eid = {A6},
        pages = {A6},
          doi = {10.1051/0004-6361/201730685},
archivePrefix = {arXiv},
       eprint = {1705.07090},
 primaryClass = {astro-ph.GA},
       adsurl = {https://ui.adsabs.harvard.edu/abs/2017A&A...602A...6S}
}

@ARTICLE{2017A&A...602A...1S,
       author = {{Smol{\v{c}}i{\'c}}, V. and {Novak}, M. and {Bondi}, M. and {Ciliegi}, P. and {Mooley}, K.~P. and {Schinnerer}, E. and {Zamorani}, G. and {Navarrete}, F. and {Bourke}, S. and {Karim}, A. and {Vardoulaki}, E. and {Leslie}, S. and {Delhaize}, J. and {Carilli}, C.~L. and {Myers}, S.~T. and {Baran}, N. and {Delvecchio}, I. and {Miettinen}, O. and {Banfield}, J. and {Balokovi{\'c}}, M. and {Bertoldi}, F. and {Capak}, P. and {Frail}, D.~A. and {Hallinan}, G. and {Hao}, H. and {Herrera Ruiz}, N. and {Horesh}, A. and {Ilbert}, O. and {Intema}, H. and {Jeli{\'c}}, V. and {Kl{\"o}ckner}, H.-R. and {Krpan}, J. and {Kulkarni}, S.~R. and {McCracken}, H. and {Laigle}, C. and {Middleberg}, E. and {Murphy}, E.~J. and {Sargent}, M. and {Scoville}, N.~Z. and {Sheth}, K.},
        title = "{The VLA-COSMOS 3 GHz Large Project: Continuum data and source catalog release}",
      journal = {\aap},
         year = 2017,
        month = jun,
       volume = {602},
          eid = {A1},
        pages = {A1},
          doi = {10.1051/0004-6361/201628704},
archivePrefix = {arXiv},
       eprint = {1703.09713},
 primaryClass = {astro-ph.GA},
       adsurl = {https://ui.adsabs.harvard.edu/abs/2017A&A...602A...1S}
}

@ARTICLE{2022ApJS..258...11W,
       author = {{Weaver}, J.~R. and {Kauffmann}, O.~B. and {Ilbert}, O. and {McCracken}, H.~J. and {Moneti}, A. and {Toft}, S. and {Brammer}, G. and {Shuntov}, M. and {Davidzon}, I. and {Hsieh}, B.~C. and {Laigle}, C. and {Anastasiou}, A. and {Jespersen}, C.~K. and {Vinther}, J. and {Capak}, P. and {Casey}, C.~M. and {McPartland}, C.~J.~R. and {Milvang-Jensen}, B. and {Mobasher}, B. and {Sanders}, D.~B. and {Zalesky}, L. and {Arnouts}, S. and {Aussel}, H. and {Dunlop}, J.~S. and {Faisst}, A. and {Franx}, M. and {Furtak}, L.~J. and {Fynbo}, J.~P.~U. and {Gould}, K.~M.~L. and {Greve}, T.~R. and {Gwyn}, S. and {Kartaltepe}, J.~S. and {Kashino}, D. and {Koekemoer}, A.~M. and {Kokorev}, V. and {Le F{\`e}vre}, O. and {Lilly}, S. and {Masters}, D. and {Magdis}, G. and {Mehta}, V. and {Peng}, Y. and {Riechers}, D.~A. and {Salvato}, M. and {Sawicki}, M. and {Scarlata}, C. and {Scoville}, N. and {Shirley}, R. and {Silverman}, J.~D. and {Sneppen}, A. and {Smolc̆i{\'c}}, V. and {Steinhardt}, C. and {Stern}, D. and {Tanaka}, M. and {Taniguchi}, Y. and {Teplitz}, H.~I. and {Vaccari}, M. and {Wang}, W.-H. and {Zamorani}, G.},
        title = "{COSMOS2020: A Panchromatic View of the Universe to z{\ensuremath{\sim}}10 from Two Complementary Catalogs}",
      journal = {\apjs},
         year = 2022,
        month = jan,
       volume = {258},
       number = {1},
          eid = {11},
        pages = {11},
          doi = {10.3847/1538-4365/ac3078},
archivePrefix = {arXiv},
       eprint = {2110.13923},
 primaryClass = {astro-ph.GA},
       adsurl = {https://ui.adsabs.harvard.edu/abs/2022ApJS..258...11W}
}

@ARTICLE{2016ApJ...819...62C,
       author = {{Civano}, F. and {Marchesi}, S. and {Comastri}, A. and {Urry}, M.~C. and {Elvis}, M. and {Cappelluti}, N. and {Puccetti}, S. and {Brusa}, M. and {Zamorani}, G. and {Hasinger}, G. and {Aldcroft}, T. and {Alexander}, D.~M. and {Allevato}, V. and {Brunner}, H. and {Capak}, P. and {Finoguenov}, A. and {Fiore}, F. and {Fruscione}, A. and {Gilli}, R. and {Glotfelty}, K. and {Griffiths}, R.~E. and {Hao}, H. and {Harrison}, F.~A. and {Jahnke}, K. and {Kartaltepe}, J. and {Karim}, A. and {LaMassa}, S.~M. and {Lanzuisi}, G. and {Miyaji}, T. and {Ranalli}, P. and {Salvato}, M. and {Sargent}, M. and {Scoville}, N.~J. and {Schawinski}, K. and {Schinnerer}, E. and {Silverman}, J. and {Smolcic}, V. and {Stern}, D. and {Toft}, S. and {Trakhtenbrot}, B. and {Treister}, E. and {Vignali}, C.},
        title = "{The Chandra Cosmos Legacy Survey: Overview and Point Source Catalog}",
      journal = {\apj},
         year = 2016,
        month = mar,
       volume = {819},
       number = {1},
          eid = {62},
        pages = {62},
          doi = {10.3847/0004-637X/819/1/62},
archivePrefix = {arXiv},
       eprint = {1601.00941},
 primaryClass = {astro-ph.GA},
       adsurl = {https://ui.adsabs.harvard.edu/abs/2016ApJ...819...62C}
}
\bibliographystyle{aasjournalv7.1}

\end{document}